\documentclass[12pt,preprint]{aastex}

\newcommand{\etal}{et al.}

\shorttitle{Stellar Populations of AGN Hosts}
\shortauthors{Draper \&  Ballantyne}

\begin{document}

\title{The Young, the Old, and the Dusty: Stellar Populations of AGN Hosts}


\author{A. R. Draper and D. R. Ballantyne}
\affil{Center for Relativistic Astrophysics, School of Physics,
  Georgia Institute of Technology, Atlanta, GA 30332}
\email{aden.draper@physics.gatech.edu}

\begin{abstract}

Studying the average properties of active galactic nuclei (AGN) host stellar populations 
is an important step in understanding the role of AGN in galaxy evolution and the 
processes which trigger and fuel AGN activity.  Here we calculate model spectral energy distributions 
(SEDs) that include emission from the AGN, the host galaxy stellar population, 
and dust enshrouded star formation.  Using the framework of cosmic X-ray background 
population synthesis modeling, the model AGN hosts are constrained using 
optical (B band) and near infrared (J band, 3.6 $\mu$m, 5.7 $\mu$m, 8.0 $\mu$m, and 24 $\mu$m) 
luminosity functions and number counts.  It is found that at $z$ $<$ 1, type 1 and type 2 AGN 
hosts have similar stellar populations, in agreement with the orientation based unified model and 
indicative of secular evolution.  At $z$ $>$ 1, type 2 AGN hosts are intrinsically different from 
type 1 AGN hosts, suggesting that the simple orientation based unified model does not hold at $z$ $>$ 1.  
Also, it is found that if Compton thick (CT) AGN evolve like less obscured type 2 AGN, then, on average, 
CT AGN hosts are similar to type 2 AGN hosts; however, if CT obscuration is connected to an evolutionary 
stage of black hole growth, then CT AGN hosts will also be in specific evolutionary stages. Multi-wavelength 
selection criteria of CT AGN are discussed.

\end{abstract}


\section{Introduction}
\label{sect:intro}

It is known that all massive galaxies have a central supermassive black 
hole (SMBH) \citep[e.g.,][]{KR95}.  Yet it is unknown why only a small fraction of these SMBHs are 
actively accreting as active galactic nuclei (AGN).  
Several theories attempt to explain the fueling mechanism of AGN: major mergers 
\citep[e.g.,][]{S88, H06}, minor mergers and gravitational instabilities 
within the host galaxy \citep[e.g.,][]{Cr03,KK04, P07,S08}, nuclear starbursts 
\citep[e.g.,][]{D07,B08}, supernova explosions within nuclear starbursts 
\citep[e.g.,][]{C09, KJ10}, and collisions of warm halo clouds with the 
nuclear region \citep{M10}.  Given the large range of observed AGN properties, 
it is likely that different fueling mechanisms come into play for different 
AGN populations.  Current observations suggest that powerful quasars are triggered 
by major mergers but moderate luminosity AGN are more likely to be triggered by 
stochastic fueling incidents \citep{B06a,H08,HH09,L10}.  However, it is not yet clear which 
fueling mechanisms are dominant in which portions of the AGN population.  

The unified model of AGN \citep{A93} explains the different observed 
levels of AGN obscuration as a simple geometric effect.  This model 
assumes that the central engines of all AGN are identical and the 
level of obscuration is dependent on the line of sight between the 
observer and the central engine.  Thus type 2 AGN are obscured by column densities $N_H$ $\gtrsim$ 10$^{22}$ cm$^{-2}$, because the 
observer is looking through the dusty torus, in contrast to type 1 AGN which 
are unobscured ($N_H$ $<$ 10$^{22}$ cm$^{-2}$) because the observer is looking down the throat of the dusty 
torus.  Since the central engines are 
identical, it is expected that, on average, the host galaxies of 
various spectral types of AGN will be similar.  However, different 
black hole fueling mechanisms are likely to lead to different relationships 
between the various spectral types of AGN.  In the major merger AGN fueling paradigm, 
the AGN is triggered while deeply embedded in gas and dust and 
thus when the AGN first turns on, it is highly obscured.  Eventually 
the radiation pressure pushes away the remaining gas and dust 
revealing an unobscured quasar \citep[e.g.,][]{P04,H06,Ri09}.  In the 
major merger paradigm, type 1 and type 2 AGN have host galaxies 
which are in different evolutionary stages.
Conversely, in the nuclear starburst fueling paradigm, type 1 and type 2 AGN hosts 
can be similar since on average the unified model can hold \citep{B08}.    
This shows that tests of the unified model can be used to explore 
which processes are viable options for AGN fueling and in which sections  
of the AGN population different mechanisms are relevant. 

Classifying the host galaxies of Compton thick (CT) AGN, AGN with $N_H$ $\gtrsim$ 
10$^{24}$ cm$^{-2}$, is of special interest.  
It is expected that a large fraction of AGN with CT levels of obscuration are 
in a young evolutionary stage characterized by rapid black hole and host bulge 
growth \citep{S88,F99,H06,F09,DB10,T10}.  Thus, 
understanding the host properties of CT AGN may offer special 
insight into the AGN triggering process and the formation of massive galaxies.  Large samples of CT AGN 
are difficult to identify because, by definition, CT AGN suffer from extreme 
levels of obscuration \citep[e.g.,][]{Ghis94}.  Therefore, previous studies of AGN host galaxies 
have not been able to study the stellar populations of CT AGN.

Galaxy optical colors are often described in terms of the red sequence and the 
blue cloud, where objects located on the red sequence are characterized by massive older 
stellar populations and objects on the blue cloud are less massive and have young stellar populations 
and current star formation.  Between the red sequence and the blue cloud is a 
lightly populated region referred to as the green valley.  Some studies find that 
AGN feedback is connected to the shut down of host star formation \citep[e.g.,][]
{Mc10, B11} and thus certain types of AGN hosts preferentially reside in the green 
valley \citep[e.g.,][]{H09,GS10,W10,SR11}.  Thus AGN activity is potentially a stage 
in galaxy evolution where feedback from the AGN could shut down star formation, 
causing the host galaxy to age from the blue cloud, across the green valley, and 
onto the red sequence \citep{F07,S07}.  However, other studies find that the colors 
of an AGN and its host are more closely related to the amount of available 
obscuring material rather than the evolutionary stage of the host stellar population 
\citep{B09,Geo09,C10,R11}.  Indeed, \citet{C10} show that AGN are intrinsically 
bimodal in color and that many AGN which appear to be on the red sequence are actually 
located in the blue cloud once dust extinction is taken into account.  Thus further 
investigation is necessary to establish the nature of AGN host galaxy stellar 
populations and the role of AGN feedback in regulating star formation.  This point 
is especially salient as several possible AGN fueling 
mechanisms include either causal or concurrent star formation.  
Some studies even show that AGN obscuration may in part be due to nuclear starburst disks 
\citep[e.g.,][]{T05, D07,B08}.  In order to understand how host galaxy processes 
affect the central SMBH and the role of the central SMBH in host galaxy evolution, 
the nature of stellar populations of AGN hosts must be well understood.

As AGN hosts exhibit size-able object-to-object variability, it 
is difficult to elucidate trends in the AGN population by fitting 
individual objects.  Thus, in order to study larger trends in AGN--host galaxy
interactions, the average stellar properties of a large 
ensemble of AGN hosts is investigated here.  To this end the stellar synthesis models of 
\citet{bc03} are used to explore the stellar populations of AGN hosts.  
As in previous studies \citep{B06,DB11}, the AGN SEDs 
are calculated using the photoionization code Cloudy \citep{F98}.  The cosmic 
X-ray background (CXRB) synthesis modeling framework is used to characterize the AGN 
population.  The model SEDs are then used to move the CXRB framework 
into other wavelength regions.  This allows the space density and evolution of AGN 
host galaxies to be determined by the most comprehensive census of AGN activity.  
By comparing these models against various observations 
in the optical through mid-infrared (mid-IR) spectral ranges, constraints are placed on 
the average stellar populations of type 1, Compton thin type 2 (here referred 
to simply as type 2), and CT AGN at various redshifts.  In Section 
\ref{sect:model} the AGN spectral model and stellar population model are described.  
Section \ref{sect:results} presents the results of the model.  In Sections 
\ref{sect:disc} and \ref{sect:sum} the results are discussed and summarized.  AB 
magnitudes are used, unless otherwise stated, and $h$ = 0.7, $\Omega_{\Lambda}$ = 
$1-\Omega_{M}$ = 0.7 is assumed as necessary.

\section{The Model}
\label{sect:model}

\subsection{Methodology}
\label{sub:method}

\begin{figure*}
\begin{center}
\includegraphics[angle=0,width=0.95\textwidth]{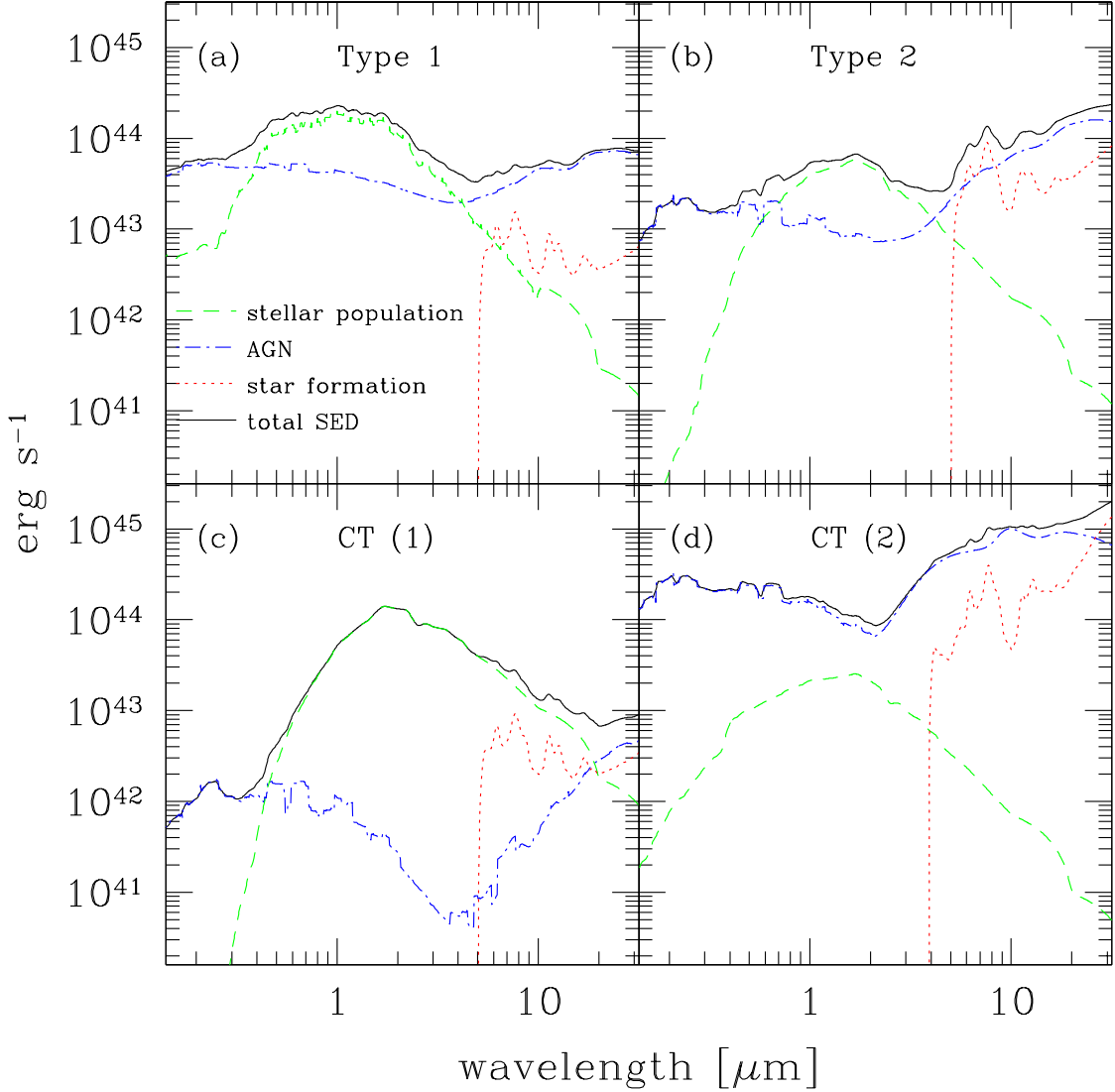}
\end{center}
\caption{Smoothed AGN and host rest frame SEDs.  Solid lines show the total AGN and host SED.  Dot-dashed lines show AGN SEDs, dashed lines show \citet{bc03} stellar population model SEDs, and dotted lines show \citet{R09} dusty star formation templates.  (a) type 1 AGN with $L_X$ = 10$^{43}$ erg s$^{-1}$, a stellar population of $M_*$ = 10$^{11}$ M$_{\odot}$ with 90$\%$ of $M_{*}$ in a 4.5 Gyr old stellar population and 10$\%$ of $M_*$ in a 2 Gyr old stellar population with $E(B-V)$ $\approx$ 0.25, and the $L_{IR}$ = 10$^{10}$ L$_{\odot}$ dusty star formation template, corresponding to SFR $\approx$ 2 M$_{\odot}$ yr$^{-1}$.  (b) type 2 AGN with $L_X$ = 10$^{43}$ erg s$^{-1}$, the same stellar population as in (a) but with $E(B-V)$ $\approx$ 0.50, and the $L_{IR}$ = 10$^{11}$ L$_{\odot}$ template, corresponding to SFR $\approx$ 17 M$_{\odot}$ yr$^{-1}$.  (c) low Eddington ratio CT AGN with $L_X$ = 10$^{42}$ erg s$^{-1}$, a 4.5 Gyr old stellar population of $M_*$ = 10$^{12}$ M$_{\odot}$ with $E(B-V)$ $\approx$ 1.0, and SFR $\approx$ 1 M$_{\odot}$ yr$^{-1}$, corresponding to the $L_{IR}$ = 10$^{9.75}$ L$_{\odot}$ template.  (d) high Eddington ratio CT AGN with $L_X$ = 10$^{44}$ erg s$^{-1}$, a 0.3 Gyr old stellar population of $M_*$ = 10$^{10}$ M$_{\odot}$ with $E(B-V)$ $\approx$ 0.50, and the $L_{IR}$ = 10$^{12}$ L$_{\odot}$ template, or SFR $\approx$ 175 M$_{\odot}$ yr$^{-1}$.}
\label{fig:sed}
\end{figure*}

In order to study the average properties of AGN host galaxy stellar populations, 
a model is created which combines knowledge of AGN space density and evolution 
from deep X-ray surveys and CXRB population synthesis models 
with stellar population studies.  A model SED is computed which includes the emission 
from the AGN, the host galaxy stellar population, and ongoing dust enshrouded star 
formation.  This AGN and host SED covers the wide spectral range of hard X-ray to mid-IR.  
The calculation of the AGN model SED is discussed in Section \ref{sub:AGN} and the details 
of the host stellar population model SED calculation are discussed in Section \ref{sub:bc03}.  To account 
for dust enshrouded star formation, the templates of \citet{R09} are used, which extend from 5 $\mu$m to 30 cm, 
and are determined by averaging empirical SEDs of local purely star forming galaxies.  
While dust enshrouded star formation may contribute to the galaxy SED at wavelengths shorter than 5 $\mu$m, 
generally the near IR AGN and host SEDs are dominated by either emission from the AGN or the \citet{bc03} 
portion of the host SED.  Examples of the rest frame AGN and host SEDs are shown 
in Figure \ref{fig:sed}, where the dot-dashed lines show AGN SEDs, the dashed lines show stellar population SEDs, 
and the dotted lines show the \citet{R09} templates.  

A Gaussian stellar mass ($M_{*}$) distribution was used with $M_{*}^{min}$ = 10$^{9.5}$ M$_{\odot}$ and $M_{*}^{max}$ = 10$^{12}$ M$_{\odot}$.  The 
average $M_{*}$ was set at 10$^{10.9}$ M$_{\odot}$ with a standard deviation of 
0.4 dex, in agreement with the sample of X-ray selected, $z$ $\sim$ 1 AGN hosts presented 
by \citet{P10}.  It is assumed that the mass-to-light ratio of the host galaxy is constant within each waveband
and thus the host galaxy flux scales linearly with the host $M_*$.  Therefore, the 
stellar population SED is calculated for $M_{*}$ = 1 M$_{\odot}$ and then re-normalized to $M_{*}$ = 
10$^{9.5}$ -- 10$^{12}$ M$_{\odot}$ before the stellar population and AGN SEDs are combined.  Creating 
host SEDs at several masses in the AGN host mass spectrum allows the consideration of AGN and host 
colors for hosts of various masses.  When considering space densities, however, it is important to 
consider not only the range of AGN host masses but also the probability that an AGN host be a certain $M_{*}$.  The space 
density calculations presented are therefore weighted sums across the mass spectrum of AGN 
hosts where the weight is determined by the Gaussian $M_{*}$ distribution described above.  Specifically, 
when calculating luminosity functions and number counts the weighted sum over the Gaussian $M_{*}$ 
distribution is used; however, when considering flux-flux plots, only the range of $M_*$ is used.
In agreement with current observational results, the $M_{*}$ 
distribution used does not evolve with redshift \citep{B11}.

A variety of broad band data from the optical through mid-IR is used to constrain the host models.  
These model constraints are described in Section \ref{sub:constrain}, and include mid-IR number counts, 
the J band AGN and host space density as a function of redshift and the absolute J band magnitude, $M_J$, and the type 1 AGN and host B band luminosity function.  We 
begin by assuming that the unified model \citep[e.g.,][]{A93} holds and therefore, since all AGN are essentially the 
same system viewed along different lines of sight, there is only one class of AGN host galaxies.

\subsection{CXRB Synthesis Model}
\label{sub:XRB}

It is known that the CXRB encodes the accretion history of SMBHs and as such is a powerful tool for 
scientific inquiry into accretion processes \citep{FB92}.  The CXRB synthesis model framework can be 
used in wavelength regions outside of the X-ray spectral region by modeling the SEDs of AGN and their 
host galaxies.  Thus the advances in CXRB synthesis modeling \citep[e.g.,][]{R99, B06a, G07, T09b, DB09} can 
be used to further the understanding of AGN at all wavelengths.  Particularly, CXRB synthesis models 
provide constraints on the distribution of $N_H$, the fraction of 
type 2 AGN, $f_2$, and the fraction of CT AGN, $f_{CT}$.  Here a simple $N_H$ 
distribution is assumed where type 1 AGN are evenly distributed among $\log N_H$ = 20.0, 20.5, 21.0, 21.5, 
type 2 AGN are evenly distributed among $\log N_H$ = 22.0, 22.5, 23.0, 23.5, and CT AGN are evenly distributed 
among $\log N_H$ = 24.0, 24.5, 25.0.  \citet{B06a} showed that this $N_H$ distribution is consistent with CXRB 
observations and provides results which are negligibly different from the observed $N_H$ distribution of 
\citet{R99} \citep{B06a,B06}.  The shape of the CXRB combined with deep X-ray surveys has also been used to 
constrain $f_2$ \citep[e.g.,][]{B06a}.  At a given $L_X$ and redshift, $f_2$ $\propto$ $(1+z)^a(\log L_X)^{-b}$ 
where $a$ = 0.4 \citep{B06a} and $b$ = 4.7.  The normalization of $f_2$ is set such that at $z$ = 0 and 
$\log L_X$ = 41.5, the type 2 to type 1 ratio is 4:1 \citep[see Section 2.2 of][]{DB09}.

Both the original and the composite model of CT AGN evolution of \citet{DB10} 
are considered.  The original, non-evolving model assumes that CT AGN are a simple extension of the 
Compton thin type 2 AGN population and that, when assuming the \citet{U03} hard X-ray luminosity function, 
$\sim$44$\%$ of all obscured AGN are CT.  In this model 
there is a population of CT AGN which are nearly as common as Compton thin type 2 AGN and which are 
in all ways similar to Compton thin type 2 AGN except for the presence of more obscuration along our line of sight.
Contrastingly, the composite model assumes that CT AGN are a population of AGN distinct 
from the Compton thin type 2 AGN population.  Simulations show that gas rich galaxy mergers 
will cause dust and gas to be funneled into the nuclear region of the galaxy, triggering star 
formation and accretion onto the central SMBH \citep[e.g.,][]{H06}.  Due to the large reservoir 
of material, the SMBH will accrete very rapidly and be very highly obscured \citep{F99, F08, F09}.  
Thus it is expected that CT AGN are high Eddington ratio sources.  \citet{DB10} found that if all 
CT AGN are rapidly accreting sources, than the local space density of CT AGN is under-predicted, and 
thus a population of low Eddington ratio CT AGN are necessary to explain the observed space density 
of CT AGN.  These low accretion rate CT AGN are likely obscured by molecular clouds within the host 
bulge.  Observational evidence of this low luminosity CT AGN population has been found in the local 
universe \citep{TW03, GM09}.  Furthermore, \citet{DB10} showed that if CT AGN have moderate Eddington 
ratios, the space density of CT AGN with $L_X$ $>$ 10$^{43}$ erg s$^{-1}$ at $z$ $\gtrsim$ 0 is greatly over-predicted.
Therefore, the evolving model of CT AGN has an Eddington ratio dependent $f_{CT}$ with 
$\sim$86$\%$ of all AGN which are accreting at $>$90$\%$ 
of their Eddington rate being CT and $\sim$60$\%$ of all AGN which are accreting at $<$1$\%$ of their 
Eddington rate being CT.  In this model CT AGN are a distinct population of AGN associated with specific stages of AGN evolution
and $f_{CT}$ evolves with both $L_X$, the AGN 2--10 keV luminosity, and redshift.  Both the evolving model and the non-evolving model 
are consistent with the CXRB and the local space density of CT AGN with $L_X$ $>$ 10$^{43}$ 
erg s$^{-1}$ \citep{DB10}.

\subsection{AGN SEDs}
\label{sub:AGN}
In order to compute the model AGN SEDs, the photoionization code 
Cloudy version C08.00 \citep{F98} is used, following the same procedure  
as in Section 2 of \citet{DB11}.  These SEDs cover the wavelength 
range of very hard X-ray through far-IR and include the transmitted 
AGN emission, the diffuse emission emitted along the line of sight 
by the obscuring material around the AGN, and the reflected emission 
off the inner face of the obscuring cloud.  The inner radius of the 
obscuring material is assumed to be $\sim$10 pc.  As in \citet{DB11}, 
the obscuring clouds are assigned neutral hydrogen densities, $n_H$, 
in agreement with typical molecular clouds; the Compton thin clouds 
have $n_H$ = 10$^4$ cm$^{-3}$ and the Compton thick clouds have 
$n_H$ = 10$^6$ cm$^{-3}$.

\subsection{Host Galaxy Stellar Population Model}
\label{sub:bc03}

The host galaxy stellar populations are modeled using GALAXEV \footnote{available at http://www.
cida.ve/~bruzual/bc2003} \citep{bc03}.  This model calculates stellar population SEDs 
over the wavelength range of 91\AA -- 160 $\mu$m.  Using an isochrone synthesis 
technique, GALAXEV can evolve stellar populations with three stellar evolution library 
options.  We assume the stellar evolution models of the 'Padova 1994' library and the 
stellar spectra of the STELIB/BaSeL 3.1 semi-empirical library.  GALXEV models can be 
computed for six different metallicities in the range $Z$ = 0.005Z$_{\odot}$ - 2.5Z$_
{\odot}$.  Both the \citet{S55} and \citet{C03} initial mass functions (IMFs) are 
available in the GALAXEV code.  Here we assume solar metallicity \citep{K07,S09} and 
the \citet{C03} IMF \citep{S09}. The sensitivity of the results to these assumptions is 
assessed in Section \ref{sect:results}.  GALAXEV 
also allows for a variety of star formation histories, including constant star formation 
rate, instantaneous bursts, and exponentially declining star formation rate.  An 
exponentially declining star formation history \citep{K07,S09} with an $e$-folding time of 0.5 Gyr is used 
here.  This star formation history allows for a simple parametrization of the 
average age of the stellar population while not requiring the entire stellar population to be a single age.
GALAXEV does not include re-radiation of energy absorbed by dust; however, the majority 
of this re-emission by cool dust occurs at wavelengths longer than those considered here.  

The stellar population SEDs are computed in four different redshift bins: $z$ $<$ 1, 1 $<$ $z$ $<$ 2, 
2 $<$ $z$ $<$ 3, and 3 $<$ $z$ $<$ 5, where $z$ is the redshift.  We assume that the host galaxy has a dominate stellar 
population (DSP) which is slightly younger than the average age of the universe in each redshift bin, 
and $M_{*}$ is given by the Gaussian distribution described in Section \ref{sub:method}.  
A younger stellar population (YSP) which accounts for $\sim$ 0.1$M_{*}$ \citep{K07,Sh09} is also included.  The age of the 
YSP is selected such that $D_n$(4000\AA) $\approx$ 1.5 
\citep{K03, S09}, using $D_n$(4000\AA) = $f_{RC}/f_{BC}$, where $f_{RC}$ is the flux in the 4000--4100 {\AA} 
wavelength range and $f_{BC}$ is the flux in the 3850--3950 {\AA} wavelength range \citep{B99}.  For the stellar 
population models used here, $D_n$(4000\AA) $\approx$ 1.5 corresponds to a stellar population age of 
$\sim$2 Gyr.  The age of the YSP and DSP in each redshift bin is summarized in Table \ref{ages}.  
At $z$ $>$ 2, the older and younger stellar populations are the same age and thus at 2 $<$ $z$ $<$ 3 the entire stellar 
population is 2 Gyr old and at 3 $<$ $z$ $<$ 5 the entire stellar population is 1 Gyr old. 
Extinction due to dust, as described by $E(B-V)$, is allowed to vary between redshift bins, but not within redshift bins.  
While in some objects the AGN emission is likely to be extincted by extended dust structures within the host galaxy 
\citep[e.g.,][]{MS10}, this is not true for all objects.  As this study focuses on modeling the average properties of 
AGN and their hosts, a detailed modeling of the complex geometry of obscuring material in individual sources is beyond the scope of this work.  Thus,  
it is assumed here that all of the extinction suffered by the AGN emission occurs within a few tens of parsecs of the central engine.  
The dust which extincts the stellar population is not allowed to further extinct the AGN emission.


\subsection{Model Constraints}
\label{sub:constrain}

A variety of data in a broad range of spectral regions was used to constrain the stellar population 
models.  The number 
counts are computed at 3.6, 5.7, 8.0 and 24 $\mu$m and compared to the 
AGN observed by {\em Spitzer Space Observatory} in the GOODS fields with $f_{2-8}$ $\gtrsim$ 1 $\times$ 10$^{-16}$ erg s$^{-1}$ cm$^{-2}$, 
where $f_{2-8}$ is the 2--8 keV flux \citep{T06}.  This X-ray flux limit is taken into consideration in the 
number counts calculated here, thus the number counts presented are computed for the same population probed 
by the observational data points.  These number counts are dominated 
by type 2 AGN and their hosts and therefore are primarily useful to constrain the type 
2 AGN host population.  The number of sources at wavelength $\lambda$ per square 
degree with flux greater than $S$, $N_{\lambda}$($>S$), is found by
\begin{equation}
N_{\lambda}(>S) = \frac{K^{deg}_{sr}c}{H_0} 
\times \int^{z_{max}}_{z_{min}} \int^{\log L_X^{max}}_{max(\log L_X^{min}, \log L_X^S)} \frac{d\Phi (L_X, z)}{d\log L_X} \frac{d_l^2}{(1+z)^2[\Omega_m(1+z)^3+\Omega_{\Lambda}]^{1/2}} d\log L_X dz,
\label{eq:counts}
\end{equation}
where $K^{deg}_{sr}$ = 3.05 $\times$ 10$^{-4}$ deg$^{2}$ sr$^{-1}$, 
$d\Phi /d\log L_X$ is the hard X-ray luminosity function of \citet{U03} 
or, when referring to the evolving model, is the evolving Eddington 
ratio space density calculated by \citet{DB10}, in units of Mpc$^{-3}$, 
$d_l$ is the luminosity distance, and $\log L_X^S$ is the 2--10 keV 
rest-frame luminosity which corresponds to the observed-frame flux $S$ 
at redshift z.    

As the number counts are dominated by type 2 AGN and their hosts, a separate data set must be used 
to constrain the type 1 AGN host stellar populations.  The most obvious choice is the 
optical luminosity function for type 1 Seyferts and QSOs and their hosts.  Using the X-ray luminosity function, 
the type 1 AGN and host B band luminosity function, $d\Phi_B/d(mag_B)$ can be calculated using 
the following equation:
\begin{equation}
\frac{d\Phi_B}{d(mag_B)} = (1.0-f_2) \times \frac{d\Phi}{d(\log L_X)}\frac{d(\log L_X)}{d(mag_B)}.
\label{eq:BLF}
\end{equation}
This luminosity function is considered in three redshift bins --- $z$ $<$ 0.4, 1.0 $<$ $z$ $<$ 1.55, and 1.55$<$ $z$ $<$ 2.1.

%
%

Further constraints were considered in order to rule out possible degeneracies 
between stellar population age and extinction due to absorption by dust.  The 
first of these additional constraints was the AGN and host J band space density as a 
function of redshift in several $M_J$ bins.  First, the AGN and host J band luminosity 
function, $d\Phi_J/d(mag_J)$, must be calculated. This is done in the same manner 
as $d\Phi_B/d(mag_B)$ above.  Therefore,  
\begin{equation}
\frac{d\Phi_J(M_J,z)}{d(mag_J)}=A\times\frac{d\Phi}{d(\log L_X)}\frac{d(\log L_X)}{d(mag_J)}.
\label{eq:Jband}
\end{equation}
The normalization constant $A$ depends on the AGN spectral type being considered.  For type 
1 AGN, $A$ = $(1.0-f_2)$, while for type 2 AGN, $A$ = $f_2$, and for CT AGN, $A$ = $f_{CT}$.
For comparison with the J band AGN and host space density measured by \citet{A11}, $d\Phi_J/d(mag_J)$ 
is then binned into six $M_J$ bins.  

Additionally, the optical colors were considered in the form of the U-B 
versus M$_B$ color-magnitude diagram (CMD).  Also, it is well documented that AGN and 
their hosts tend to have X-ray to optical flux ratios of 0$\pm$1 \citep[e.g.][]
{A01, B01, R10}, with CT AGN hosts generally falling below this ratio.  Here the ratio between
the soft X-ray flux, $f_{0.5-2}$, and the R band flux, $f_R$, $\log(f_{0.5-2}/
f_R)$ = 0$\pm$1 is used to ensure that the absolute optical fluxes are in agreement 
with observations.

\subsection{Procedure}
\label{sub:proc}

With all the necessary ingredients in place, we begin by assuming that the unified model holds, 
and therefore, on average, type 1, type 2, and CT AGN hosts are identical.  The AGN space density, 
$N_H$ distribution, type 1/type 2 AGN ratio, and $f_{CT}$ are set by the CXRB model used.  The age of the stellar 
populations is assigned such that the dominant stellar population age is slightly less than the 
mean Hubble time in each redshift bin and a younger stellar population is assigned the age 
corresponding to the average observed $D_n$(4000 {\AA}) for AGN host galaxies.  The stellar population ages are 
summarized in Table \ref{ages}.  The host $M_{*}$ 
distribution is Gaussian within the observed AGN host $M_{*}$ range.  The only free parameters are
the host galaxy dust enshrouded star formation rate and the extinction due to 
dust.  These two free parameters are used to fit the near and mid-IR AGN and host number counts.  
The dust enshrouded star formation rate is determined by cycling through the \citet{R09} templates in order of 
increasing infrared luminosity in steps corresponding to a change in SFR of $\sim$1 M$_{\odot}$ 
yr$^{-1}$, until the 24 $\mu$m number counts are over-estimated at all flux levels.  The template of the 
highest luminosity to not over-estimate the mid-IR number counts is selected as the best fit template.  
The dust extinction in the \citet{bc03} models is set using the total effective V band optical depth 
obscuring young stars, $\tau_{V}$.  SEDs are computed with $\tau_V$ = 0.0, 1.0, 2.0, 5.0, 7.5, 10, 15, 20, and 25.  
This corresponds to $E(B-V)$ = 0.0 - $\sim$1.0, with the exact values of $E(B-V)$ depending on the age of the 
stellar population.  The $E(B-V)$ with a reduced $\chi^{2}$, $\chi_{red}^2$, closest to 1.0, with respect to the 
35 number counts data points at 3.6 $\mu$m, 5.7 $\mu$m, 8.0 $\mu$m, and 24 $\mu$m, is selected as the best fit model.
The resulting model is then compared to the suite of optical and near IR observations described in Section \ref{sub:constrain} in order to evaluate the 
appropriateness of the model fit to observations.  In this manner the average AGN host galaxy properties are 
elucidated without the complications which arise from object-to-object variability.  Next, the assumption of the unified model 
is tested by considering the specific constraints for different spectral types.  Multiple star 
formation scenarios and the evolution of $f_2$ are also considered.

\section{Results}
\label{sect:results}

\subsection{Unified Model of AGN Hosts}
\label{sub:unified}

In this section, it is assumed that the unified model holds and therefore type 1, type 2, and CT AGN have, on average, identical host galaxies.  
First, the maximum dust enshrouded star formation rate (SFR) allowed by the 
near and mid-IR number counts for X-ray selected AGN observed by 
{\em Spitzer} in the GOODS fields, as reported by \citet{T06}, is determined.
Figure \ref{fig:sed} shows that the mid-IR emission is due primarily to the AGN and dust enshrouded 
star formation; therefore, fitting the 24 $\mu$m number counts can be used to set an upper limit on 
the average AGN host dust enshrouded SFR.
In order to not over-estimate the AGN and host mid-IR number counts, the average AGN host must have a  
SFR $\lesssim$ 2 M$_{\odot}$ yr$^{-1}$.  This SFR is consistent with the 
findings of \citet{B06}, who found that an AGN host SFR $\approx$ 1 M$_{\odot}$ yr$^{-1}$ provides 
a good fit to {\em Spitzer}'s measurement of the AGN contribution to  mid-IR portion of the cosmic 
infrared background, and is similar to the SFR of local 
normal spiral galaxies \citep[e.g.,][]{L09}.  According to the \citet{K98} 
relation, this SFR corresponds to an infrared 
star formation luminosity $L_{IR}$ $\approx$ 10$^{10}$ L$_{\odot}$.  

Next, the dust content of the 
average AGN host galaxy is fixed by fitting the near IR AGN and host number counts.  It is found 
that if the average AGN host has $E(B-V)$ $\approx$ 0.5, the near IR number 
counts are under-predicted.  However, if the average AGN host has $E(B-V)$ $\approx$ 
0.4, the near and mid-IR number counts are over-predicted.  As it is known that higher 
redshift galaxies tend to be dustier than local galaxies \citep[e.g.,][]{D03,Santini10}, it is assumed that $z$ 
$<$ 1 AGN hosts contain less dust than $z$ $>$ 1 AGN hosts.  If the average AGN host 
at $z$ $<$ 1 has $E(B-V)$ $\approx$ 0.25 and the average $z$ $>$ 1 AGN host has 
$E(B-V)$ $\approx$ 0.5, the predicted number counts are in good agreement with the near 
and mid-IR number counts observations, with $\chi_{red}^2$ = 1.2.  
If the average $z$ $<$ 1 AGN host has $E(B-V)$ $<$ 0.25, 
the mid-IR number counts are over-predicted.  In contrast, the model constraints are not very sensitive 
to the age of the older, dominant stellar population, which can be changed by $\sim$ 1 Gyr with only 
minor effects.  The age of the younger stellar mass population is constrained to within $\sim$0.5 Gyr. 

\begin{figure*}
\begin{center}
\includegraphics[angle=0,width=0.95\textwidth]{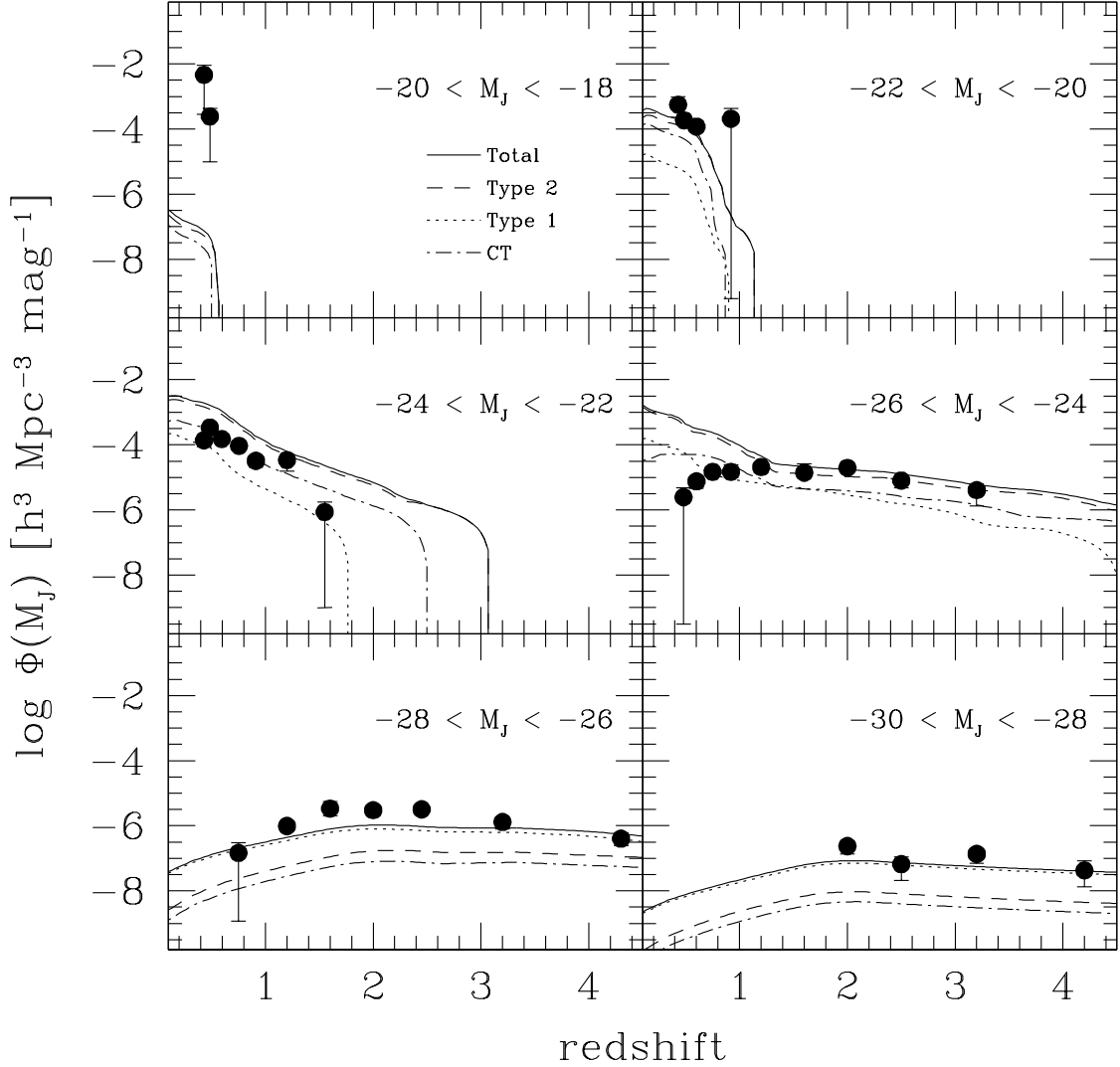}
\end{center}
\caption{J band space density for AGN and hosts for the unified AGN host model.  The black lines show the AGN and host J band space density for the model host galaxies described in row 2 of Table \ref{work}.  The solid lines show the total AGN and host J band space density while the dotted lines show the space density for type 1 AGN and hosts, dashed lines show the space density for type 2 AGN and hosts, and dot-dashed lines show the CT AGN and host space density.  Data from the mid-IR and X-ray selected AGN sample of \citet{A11} is also shown.  The obvious discrepancy at $z$ $<$ 0.75 between the model presented here and the observations reported by \citet{A11} are primarily due to a selection criterion used by \citet{A11} for optically extended sources at $z$ $<$ 0.75 which cannot be replicated here as the models do not contain information of the spatial extent of the host galaxies.}
\label{fig:jun}
\end{figure*}

This unified model of AGN hosts is also in good agreement with the other model constraints.  
The J band AGN and host space density is in decent agreement with observations, as shown in Figure \ref{fig:jun}.  
The data points in Figure 
\ref{fig:jun} show the results of \citet{A11}.  At $z$ $<$ 0.75 the IR and X-ray 
selected AGN sample of \citet{A11} was chosen using a specific selection criterion for 
optically extended sources.  As the models used here do not account for the spatial 
extent of the AGN host galaxies, this exact selection criterion cannot be replicated.  
This gives rise to an obvious discrepancy between the model predictions presented 
here and the observed J band space density reported by \citet{A11}.  Thus at $z$ $<$ 0.75 
we over-predict the AGN and host space density in magnitude bins of medium brightness and under-predict the fainter 
magnitude bins.  Type 2 AGN hosts dominate the J band space density at $M_J$ $>$ -26 for $z$ $\gtrsim$ 0.75, with 
type 1 AGN hosts dominating the brightest magnitude bins.  

\begin{figure*}
\begin{center}
\includegraphics[angle=0,width=0.95\textwidth]{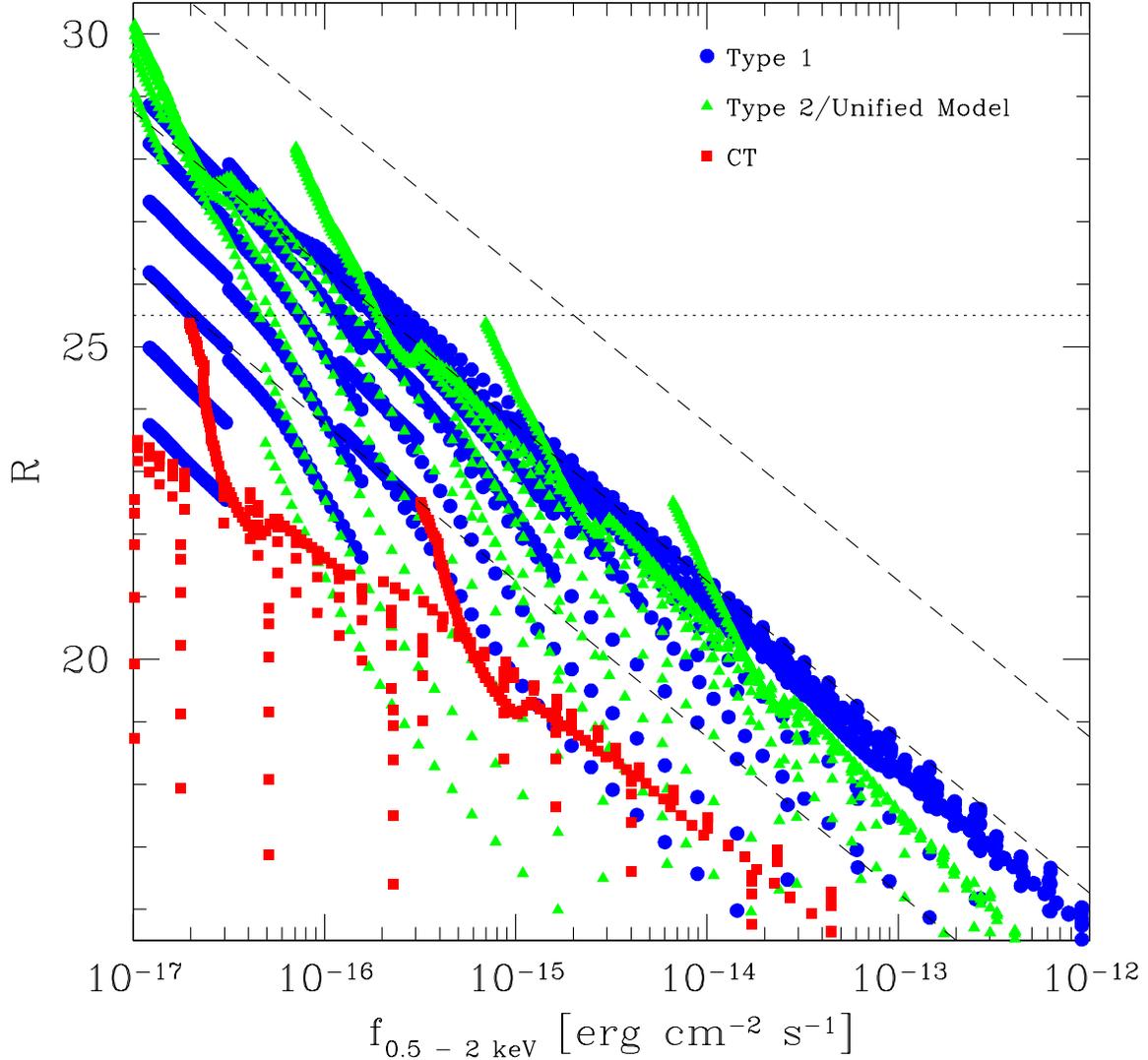}
\end{center}
\caption{Soft X-ray flux versus R band magnitude for AGN and hosts for the non-evolving model.  AGN and hosts are shown for $L_X$ $<$ 10$^{45}$ erg s$^{-1}$, $z$ $<$ 3, and $M_{*}$ = 10$^{9.5}$, 10$^{10}$, 10$^{10.5}$, 10$^{11}$, 10$^{11.5}$, and 10$^{12}$ M$_{\odot}$.  The green triangles show the unified AGN hosts model and type 2 AGN and hosts, blue circles show type 1 AGN and hosts, and red squares show non-evolving model CT AGN and hosts.  The model host galaxies shown here are described by the parameters in the first three rows of Table \ref{work}.  The dashed lines show the empirical relationship $\log(f_{0.5-2keV}/f_R)$ = 0$\pm$1.  The horizontal dotted line marks $R$ = 25.5, above which the source is considered an optically faint X-ray AGN.}
\label{fig:fxtororig}
\end{figure*}

According to Figure 6 of \citet{P10}, type 1 and type 2 AGN tend to be found in the 0.55 
$<$ U-B $<$ 1.4 and -23.5 $<$ $M_B$ $<$ -19 region of the CMD.  This region of the CMD is 
well populated by the model adopted here as many combinations of AGN $L_X$ and host galaxy 
$M_{*}$ fit this criteria.  Furthermore, the AGN and hosts fit the expected 
$\log(f_{0.5-2}/f_R)$ ratio, as shown by the green triangles in Figure \ref{fig:fxtororig}.  
Thus the unified model of AGN hosts explains the average trends of AGN and host observations 
in the optical through mid-IR spectral regions.

In order to understand how well constrained these findings are, the same procedure is 
followed using a supersolar metallicity, $Z$ = 2.5$Z_{\odot}$, a subsolar metallicity, 
$Z$ = 0.2$Z_{\odot}$, and the \citet{S55} IMF.  In all of these scenarios, it is found 
that the stellar ages described in Section \ref{sub:bc03} are able to fit the model 
constraints well; however different $E(B-V)$ values are necessary.  The columns labeled "Unified/type 2 AGN host" in Table \ref{models} 
summarize the $E(B-V)$ values for the different models considered.  As the \citet{R09} templates are 
independent of the GALAXEV stellar population models, the average AGN host dust enshrouded SFR is 
not affected by changing the metallicity or IMF of the host galaxy stellar population.

Overall, the near and mid-IR AGN and host number counts, as well as the J band space density, 
optical colors, and X-ray to optical ratio of AGN and their hosts are in good agreement with 
observations if type 1, type 2, and CT AGN have, on average, similar host galaxies.  
Also, the near and mid-IR number counts require AGN hosts at $z$ $>$ 1 be dustier than AGN hosts at $z$ $<$ 1.
We now investigate if this finding holds when spectral type specific model constraints are considered.

\subsection{Type 1 AGN Hosts}
\label{sub:t1}

\begin{figure*}
\begin{center}
\includegraphics[angle=0,width=0.95\textwidth]{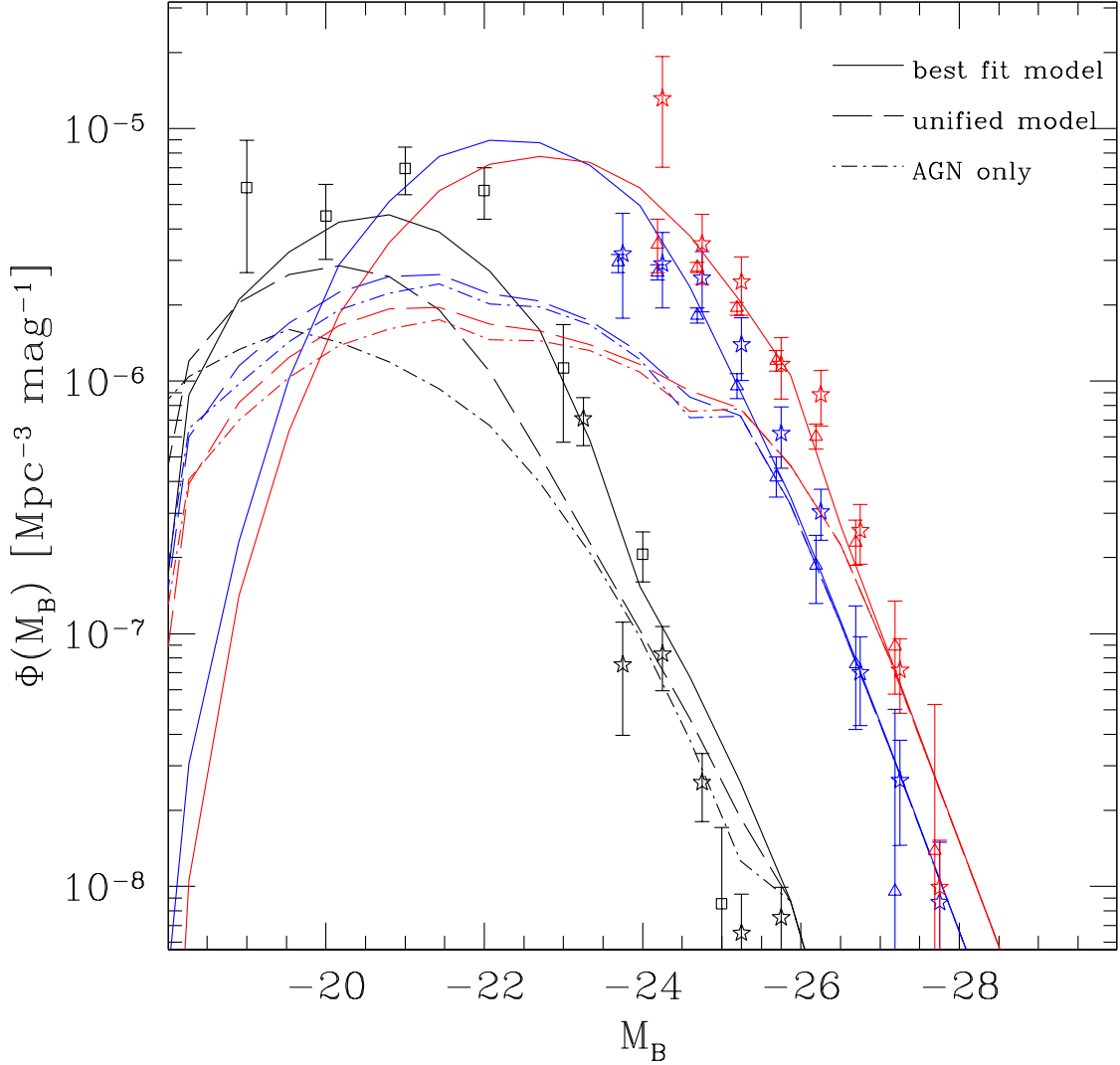}
\end{center}
\caption{Type 1 AGN and host B band luminosity function.  Different colors show the luminosity function in different redshift bins.  Black lines and data refer to $z$ $<$ 0.4, blue lines and data refer to 1.0 $<$ $z$ $<$ 1.55, and red lines and data refer to 1.55 $<$ $z$ $<$ 2.1.  The blue and red dashed lines show the luminosity function for the unified AGN host model as summarized in row one of Table \ref{work}.  The black dashed line shows the dustiest average AGN host at $z$ $<$ 1 allowed by the unified AGN host model, $E(B-V)$ $\approx$ 0.4.  The solid lines show the type 1 AGN host best fit model as summarized in row 2 of Table \ref{work}.  The AGN contribution to the luminosity function is shown as the dot-dashed lines.  Data points show various type 1 AGN and host B band luminosity functions from the literature: squares are from \citet{DC96}, triangles are from \citet{C04}, and stars are the $q_0$ = 0.5 luminosity function from \citet{HS90} converted to the cosmology used here.}
\label{fig:bband}
\end{figure*}

In order to study the host galaxies of type 1 AGN, we begin by calculating the type 1 AGN 
contribution to the type 1 AGN and host B band luminosity function.  The predicted AGN only B band luminosity 
function is shown as the dot-dashed lines in Figure \ref{fig:bband}, where black lines and 
data refer to $z$ $<$ 0.4, blue lines and data refer to 1.0 $<$ $z$ $<$ 1.55, and red lines 
and data refer to 1.55 $<$ $z$ $<$ 2.1.  The $\chi_{red}^2$ takes into account the 50 B band 
type 1 AGN and host B band luminosity function data points shown in Figure \ref{fig:bband}.  
When only the AGN contribution to the B band luminosity function is considered, 
$\chi_{red}^2$ = 14.  It is clear that except for at the very brightest 
magnitudes, a contribution from the host galaxy is necessary in order to fit the observed type 1 AGN 
and host B band luminosity function.

If the same stellar population described in Section \ref{sub:unified} is used to 
calculate the type 1 AGN and host B band luminosity function, the predicted $z$ $>$ 1 
luminosity functions are not in agreement with observations.  The unified model gives a 
$\chi_{red}^2$ = 13 fit to the B band luminosity function.  This is shown by 
the dashed lines in Figure \ref{fig:bband}.  For the $z$ $<$ 0.4 type 1 AGN and 
hosts, the lower limit $E(B-V)$ 
of the unified AGN host model B band luminosity 
function (solid black line in Figure \ref{fig:bband}) is in decent agreement with 
observations.  As the type 2 AGN hosts are constrained by the IR AGN and host number 
counts as described in Section \ref{sub:unified}, this suggests that at $z$ $<$ 1, AGN hosts can be described by the 
unified AGN host model, but at $z$ $>$ 1 the unified AGN host model cannot properly 
predict the type 1 AGN and host B band luminosity function.  Thus at $z$ $>$ 1, type 1 and 
type 2 AGN must be hosted by two distinct galaxy populations.  The lowest $\chi_{red}^2$ with respect to the B band 
luminosity function is found when type 1 AGN hosts have $E(B-V)$ $\approx$ 0.05 at $z$ $>$ 1.  For this model 
$\chi_{red}^2$ = 4.3, which is not formally a good fit to the data points, but is dominated by the over-prediction 
of the two faintest \citet{C04} data points at both 1.0 $<$ $z$ $<$ 1.55 and 1.55 $<$ $z$ $<$ 2.1.  If these points 
are not included $\chi_{red}^2$ = 2.5.  As these data 
points refer to observations of faint AGN at high redshift, it is likely that the observational sample is not complete 
at these fluxes.  Thus, at $z$ $<$ 1 type 1 and type 2 AGN 
hosts are similarly dusty.  However, at $z$ $>$ 1, type 1 AGN hosts are less dusty than type 2 AGN hosts.  Also, 
on average, it appears that type 1 and type 2 AGN hosts have similarly aged stellar populations, 
at least at $z$ $<$ 2, where the model can be well constrained. 

\begin{figure*}
\begin{center}
\includegraphics[angle=0,width=0.95\textwidth]{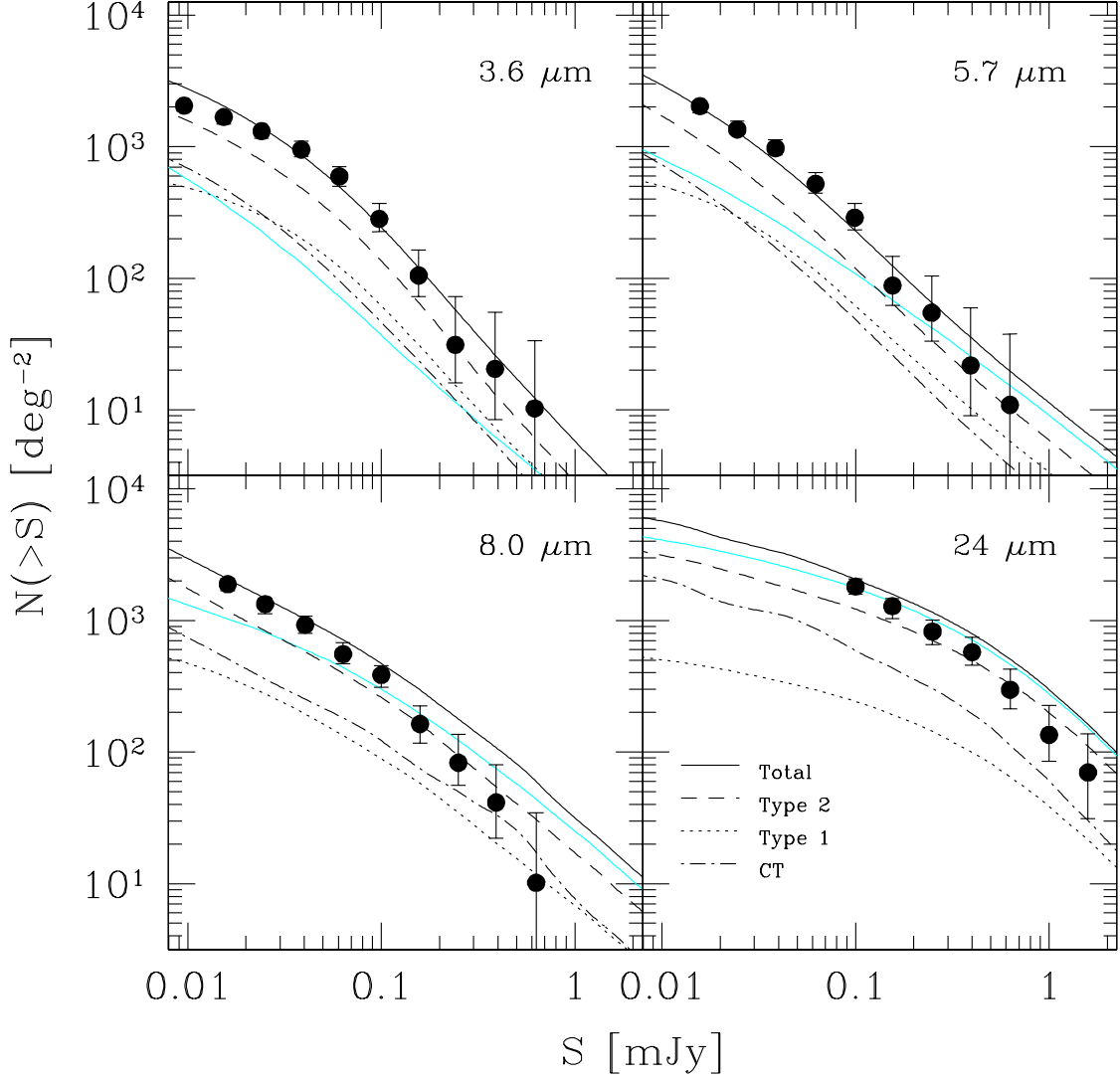}
\end{center}
\caption{Near and mid-IR number counts for AGN and hosts for the non-evolving model with $f_{2-8}$ $>$ 1 $\times$ 10$^{-16}$ erg s$^{-1}$ cm$^{-2}$.  The cyan lines show the predicted number counts for AGN alone.  The black lines show the AGN and host number counts for the model host galaxies described by the first three rows of Table \ref{work}.  The solid lines show the total AGN and host number counts while the dotted lines show the counts for type 1 AGN and hosts, dashed lines show the counts for type 2 AGN and hosts, and dot-dashed lines show the  CT AGN and host counts.  Data is from {\em Spitzer} observations of X-ray selected AGN in the GOODS fields \citep{T06}.}
\label{fig:ctsorig}
\end{figure*}

\begin{figure*}
\begin{center}
\includegraphics[angle=0,width=0.95\textwidth]{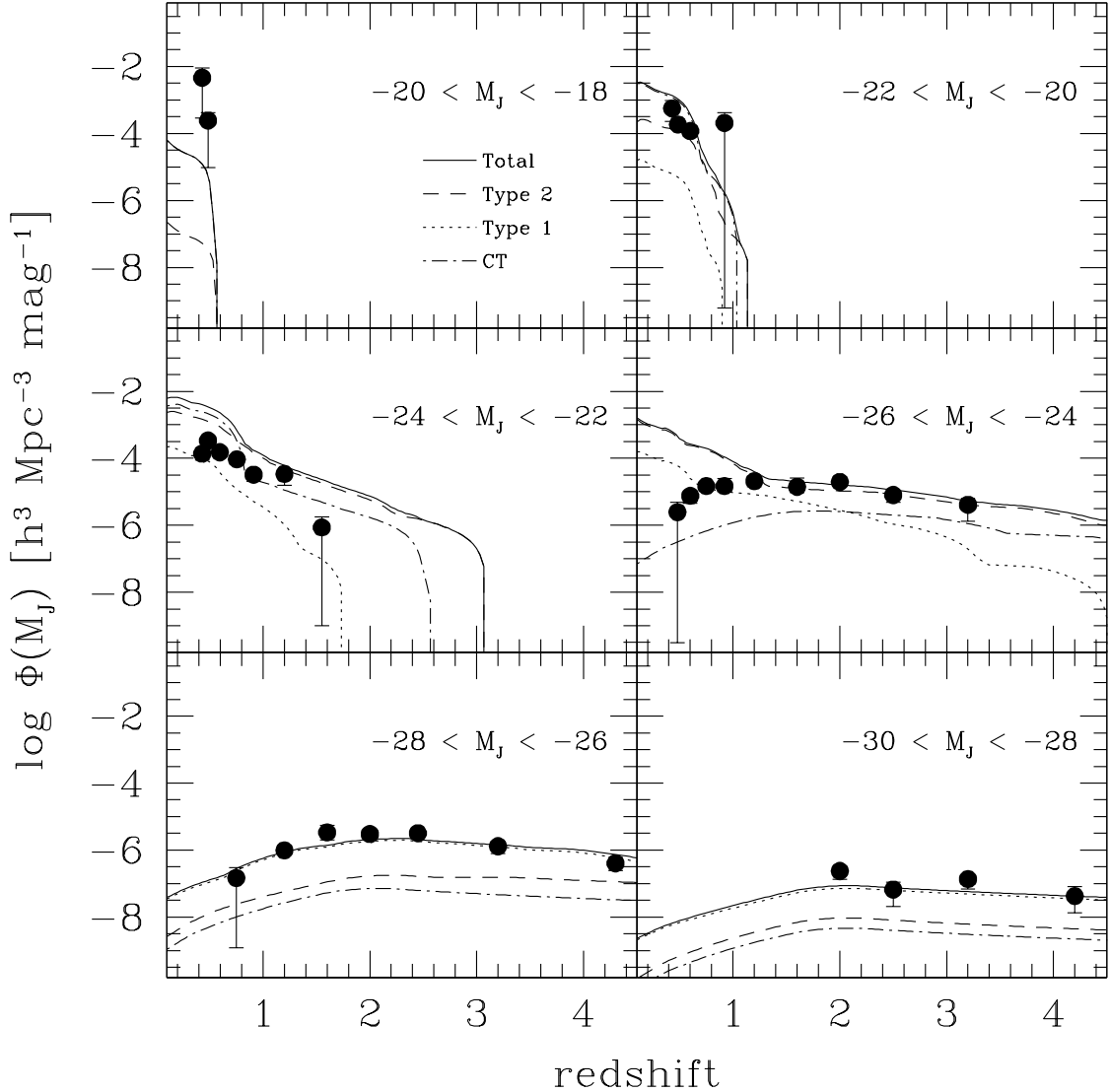}
\end{center}
\caption{J band space density for AGN and hosts for the non-evolving model.  The black lines show the AGN and host J band space density for the model host galaxies described in the first three rows of Table \ref{work}.  The solid lines show the total AGN and host J band space density while the dotted lines show the space density for type 1 AGN and hosts, dashed lines show the space density for type 2 AGN and hosts, and dot-dashed lines show the CT AGN and host space density.  Data points are the same as in Figure \ref{fig:jun}.}
\label{fig:jorig}
\end{figure*}

The type 1 AGN host stellar population described above is in good agreement 
with the other model constraints considered.  Figure \ref{fig:ctsorig} shows 
the number counts for type 1 AGN and their hosts as dotted lines.  This model has a 
$\chi_{red}^2$ = 1.3 with respect to the observed number counts.  For the unified model, 
a $\chi_{red}^2$ = 1.2 was found with respect to the IR number counts; thus reducing the dust 
extinction in type 1 AGN hosts  at $z$ $>$ 1 has little effect on the total IR number counts.  Type 1 
AGN and hosts make a considerable contribution to the shorter wavelength 
number counts, but at mid-IR wavelengths they make a minimal contribution.  
Again shown as dotted lines, Figure \ref{fig:jorig} demonstrates that type 
1 AGN and hosts dominate the J band space density for $M_J$ $<$ -26 and 
make a significant contribution in the -26 $<$ $M_J$ $<$ -24 magnitude bin.  
However, at fainter absolute magnitudes, the type 1 AGN hosts make only a 
nominal contribution to the J band AGN and host space density.  

The region of the CMD where type 1 AGN and 
their hosts tend to be found, as shown by the upside down triangles in Figure 
6a of \citet{P10}, is -23.5 $<$ $M_B$ $<$ -19 and 0.55 $<$ U-B 
$<$ 1.4.  This region of the CMD is well populated by a variety of AGN $L_X$ and host $M_*$ model
combinations.  The majority of type 1 AGN and host model SEDs used here 
are in agreement with the $\log(f_{0.5-2}/f_R)$ ratio, as exhibited 
by the blue circles in Figure \ref{fig:fxtororig}.  At fainter soft X-ray 
fluxes it appears that the type 1 AGN and hosts fall below the expected 
ratio.  These sources are located at $z$ $>$ 2.0 where the observed R band flux 
measures rest frame UV emission.  Due to the observed frame R band filter redshifting 
out of optical wavelengths, it is unclear whether the $\log(f_{0.5-2}/f_R)$ = 0 $\pm$ 1 
relation should hold at high redshift.  These high redshift type 1 AGN and hosts can be 
brought into agreement with $\log(f_{0.5-2}/f_R)$ = 0 $\pm$ 1 by increasing the extinction 
due to dust in the host stellar population or by increasing the age of the host stellar 
population.  Thus if the locally observed $\log(f_{0.5-2}/f_R)$ ratio holds at high redshift, 
this suggests that at the peak of quasar activity, type 1 AGN hosts 
either had older stellar populations than type 2 AGN hosts at the same redshift or were 
dustier than 1.0 $<$ $z$ $<$ 2.0 type 1 AGN hosts.  If the stellar populations of type 
1 AGN hosts at $z$ $>$ 2 are older than the stellar populations of type 
2 AGN hosts at the same epoch, this implies an evolutionary scenario where recently 
triggered AGN are obscured and then blow out the obscuring gas and dust in order to reveal 
an unobscured AGN and an older host.  It is also possible that, since the fraction of gas rich galaxies 
at $z$ $>$ 2 is larger than at 1 $<$ $z$ $<$ 2 \citep{Dahl07}, type 1 AGN hosts located at 
$z$ $>$ 2 contain more dust than type 1 AGN hosts located at 1 $<$ $z$ $<$ 2.  Both of these 
possibilities are consistent 
with quasars being triggered by major mergers, which is expected to be the dominate quasar fueling 
mechanism at 2 $<$ $z$ $<$ 3. 

When considering the subsolar metallicity, supersolar metallicity and \citet{S55} IMF models, 
the values for $E(B-V)$ which supply the best fit for the type 1 AGN and host B band luminosity 
function change, but the general trend that type 1 and type 2 AGN host 
stellar populations are similar at $z$ $<$ 1 and different at $z$ $>$ 1 remains.  The $E(B-V)$ 
values for type 1 AGN hosts in the different models are shown in the columns labeled "Type 1 AGN host" in Table \ref{models}.  In all models 
the type 1 AGN hosts are similar in dust content to the type 2 AGN at $z$ $<$ 1 and the type 1 AGN 
hosts are less dusty than the type 2 AGN hosts at $z$ $>$ 1.  Thus, regardless of the metallicity or 
IMF, type 1 and type 2 AGN hosts are similar at $z$ $<$ 1 and intrinsically different at $z$ $>$ 1.

\subsection{Star Formation in AGN Hosts}
\label{sub:sf}

The emission due to obscured star formation is taken into account using the 
\citet{R09} star formation templates.  The mid-IR AGN and host number 
counts suggest that the average AGN host has a dust obscured SFR $\approx$ 2 
M$_{\odot}$ yr$^{-1}$, the same SFR as normal local spiral galaxies 
\citep[e.g.,][]{L09}.  However, recent 
studies suggest that a larger, but still modest, AGN host SFR is expected 
\citep[e.g.,][]{L10, M11}.  Furthermore, studies also find that the AGN host SFR
 tends to increase with redshift \citep{L10}, AGN luminosity \citep{Th09}, or 
both \citep{SH09,S10}.  As several AGN fueling mechanisms, such as the starburst disk model of \citet{B08}, require processes 
related to star formation, it is important to consider if it is possible for 
AGN hosts to have average SFR $\gtrsim$ 2 M$_{\odot}$ yr$^{-1}$.  Thus several SFR evolutions are considered here.

The \citet{Wil10} SFR evolution depends on both AGN $L_X$ and redshift, finding 
\begin{equation}
SFR \propto \sqrt{L_X/10^{43}}(1.0+z)^{1.6}.
\label{eq:sfr}
\end{equation}
When the \citet{Wil10} SFR evolution is used, the average $z$ $<$ 1 SFR must 
still be $\lesssim$ 2 M$_{\odot}$ yr$^{-1}$ in order to not over-predict the 
mid-IR number counts.  However samples of AGN hosts with average SFRs an order 
of magnitude higher than this ($\sim$ 18-41 M$_{\odot}$ yr$^{-1}$) have been observed \citep{L10, M11, S11}.  In order 
to explain these observations, a population of enhanced star formation sources 
are considered.  The normalization factor for the proportionality in equation 
\ref{eq:sfr} is set such that the average SFR for AGN hosts at $z$ $<$ 1 is $\sim$20 
M$_{\odot}$ yr$^{-1}$.  In order to 
not over-estimate the faint end of the 24 $\mu$m number counts, enhanced star 
formation sources can account for at most $\sim$15$\%$ of the AGN population.

Similar fractions of enhanced star formation sources are found using other SFR 
evolutions.  \citet{S10} used {\em Herschel Space Observatory} observations of type 
1 SDSS selected quasars to study the evolution of SFR in quasar hosts, finding that 
AGN host SFR displays strong luminosity dependent evolution with redshift.  This leads to a SFR evolution of the form
\begin{equation}
SFR \propto (1.0+z)^{\alpha},
\label{eq:sfrs}
\end{equation}
where $\alpha$ $\approx$ -1.9$I_{AB}$ - 42, where $I_{AB}$ is the absolute I band magnitude of the quasar.  If the 
average $z$ $<$ 1 SFR is set at $\sim$20 M$_{\odot}$ yr$^{-1}$, enhanced star formation 
sources can at most be $\sim$5$\%$ of the AGN population.

If AGN host SFR does not evolve with luminosity, and instead only evolves with redshift, then 
the average AGN host SFR can be considerably higher.  If the average AGN host SFR redshift evolution
found by \citet{S10} for moderate luminosity AGN, SFR $\propto$ $(1.0+z)^{2.3}$, is applied to all
AGN hosts, an average $z$ $<$ 1 SFR $\approx$ 
16 M$_{\odot}$ yr$^{-1}$ for both type 1 and type 2 AGN is in good agreement with the 
mid-IR AGN and host number counts, with $\chi_{red}^2$ = 1.3.  This average SFR corresponds to SFR($z$=0.0) 
= 0.5 M$_{\odot}$ yr$^{-1}$, which is in excellent agreement with \citet{K06} who, using a 
sample of local type 1 SDSS quasars, found that the average local AGN host SFR $\approx$ 0.5 M$_{\odot}$ yr$^{-1}$.  
Furthermore, in the redshift only SFR evolution, for an average $z$ $<$ 1 SFR $\approx$ 20 M$_{\odot}$ yr$^{-1}$, 
$\sim$80$\%$ of AGN hosts can be enhanced star formation sources.

Thus, if the AGN host SFR evolution is dependent on luminosity, either through luminosity evolution 
or luminosity dependent redshift evolution, enhanced star formation sources are $\sim$5-15$\%$ of the 
AGN population, but if AGN host SFR only evolves with redshift, the average AGN host SFR can be up to
$\sim$16 M$_{\odot}$ yr$^{-1}$.  This finding is fully consistent with the 15 $\mu$m AGN luminosity 
function at $z$ $\sim$ 0.7 observed by \citet{Fu10}.  Furthermore, \citet{L10} find that for moderate luminosity AGN, the SFR does 
not evolve with $L_X$ and that only at the highest quasar luminosities does the AGN host SFR seem 
to depend on the AGN luminosity.  Therefore, the fraction of enhanced star formation sources found 
here is a lower limit to the true fraction of enhanced star formation sources. 



\subsection{CT AGN Hosts}
\label{sub:thick}

Observational and theoretical evidence suggests 
that CT levels of obscuration of an AGN may be due to an 
evolutionary stage where the SMBH and host bulge are 
both in a phase of rapid growth \citep[e.g.][]{S88,F99,P04,B08,F09,DB10, NR11}.  
According to galaxy merger simulations, gas rich mergers will ignite a 
burst of star formation and rapid black hole growth \citep[e.g.][]{H06}.  
In this scenario, it is expected that CT AGN hosts would be a 
subset of the ultraluminous infrared galaxy (ULIRG) population, characterized 
by $L_{IR}$ $>$ 10$^{12}$ L$_{\odot}$, where $L_{IR}$ is the 8--1000 $\mu$m 
luminosity.  In order to test this scenario, we attempt to model the CT 
AGN hosts with stellar populations similar to those found in ULIRGs 
hosting AGNs. \citet{RZ10} find that ULIRGs hosting AGNs have an average 
stellar population age of $\sim$0.3 Gyr which dominates the stellar mass 
of the galaxy with an average $M_*$ $\approx$ 10$^{10.8}$ M$_{\odot}$ and 
standard deviation of $\sim$0.35 dex and an average $E(B-V)$ $\approx$ 0.6.  
This scenario is tested using both the non-evolving and evolving 
models described in Section \ref{sub:XRB}.

Due to the observational challenges of identifying CT AGN, there is no data 
set specific to the optical or near infrared properties of CT AGN hosts.  
Thus the near and mid-IR AGN number counts measured from an X-ray flux limited 
sample of AGN, with $f_{2-8}$ $\gtrsim$ 1 $\times$ 10$^{-16}$ erg s$^{-1}$ cm$^{-2}$ \citep{T06}, 
are used to place limits on the average CT AGN host 
galaxy.  The X-ray flux limit of the {\em Spitzer} GOODS AGN sample is taken 
into account in the number counts calculation.  Thus, despite the fact that a large 
fraction of CT AGN are missed by deep X-ray surveys \citep{H08}, the dot-dashed lines in 
Figures \ref{fig:ctsorig} and \ref{fig:ctscomp} show the CT AGN which would 
have been selected by the \citet{T06} selection criteria.  
The J band space density, X-ray to optical flux ratio, and optical 
colors are also used to ensure the model CT AGN hosts are consistent with 
observations of the AGN population as a whole.

\subsubsection{Non-evolving Model}
\label{sub:orig}

The non-evolving model of \citet{DB10} assumes that CT AGN evolve 
like less obscured type 2 AGN.  In order to match the peak of the 
CXRB at $\sim$30 keV, it is required that $\sim$44$\%$ of obscured 
AGN are CT.  When the CT AGN host stellar populations 
are modeled in agreement with the findings of \citet{RZ10}, the IR number 
counts are greatly over-predicted.  This suggests that on average, CT
 AGN in the non-evolving model cannot be hosted by ULIRGs.  In fact, when using the 
non-evolving model, the average CT AGN host cannot even be a luminous 
infrared galaxy (LIRG), which is characterized by $L_{IR}$ $>$ 10$^{11}$ 
L$_{\odot}$.  In order for the near and mid-IR number counts to not be 
over-predicted at the faint end, the average non-evolving model CT AGN 
host must have $L_{IR}$ $\lesssim$ 10$^{10.75}$ L$_{\odot}$, corresponding to 
a star formation rate of $\lesssim$ 10 M$_{\odot}$ yr$^{-1}$, according to the 
\citet{K98} relation.  The youngest the non-evolving model CT AGN host 
average stellar population can be, and not over-predict the mid-IR number counts, is $\sim$ 
1 Gyr old which requires $E(B-V)$ $\approx$ 1.0.  The 
contribution of the non-evolving model CT AGN and their hosts to the near and 
mid-IR number counts is shown as the dot-dashed line in Figure \ref{fig:ctsorig}.  

In the non-evolving model, CT AGN dominate the low redshift, faint $M_{J}$ 
region of the J band space density.  Figure \ref{fig:jorig} shows the CT 
AGN and host contribution to the J band space density as the dot-dashed 
lines. As expected by the higher dust content, CT AGN hosts occupy 
a region of the CMD which is on average slightly redder and fainter than the 
type 1 and type 2 AGN hosts.  The CT AGN tend to lie 
below the $\log(f_{0.5-2}/f_R)$ ratio, as the majority of the soft X-ray 
flux of CT sources is absorbed by the CT obscuring material.   Figure \ref{fig:fxtororig} 
shows the $\log(f_{0.5-2}/f_R)$ ratio of CT AGN and their hosts as red squares.  Changing 
the metallicity or IMF used for the host galaxy stellar population does not change these results.

When the 
enhanced star formation sources are considered, the CT AGN hosts must 
be similar to the type 2 AGN hosts, regardless of host galaxy stellar 
population metallicity and IMF.  Similarly, when the redshift evolution of 
AGN host SFR is included, the CT AGN hosts must be similar to the type 2 
AGN hosts.  Thus, if $f_{CT}$ does not
evolve with AGN Eddington ratio, then CT AGN and their hosts are expected 
to be a simple extension of the less obscured
type 2 AGN population.  This would require that AGN triggered by mergers 
be a small minority of the quasar population.

\subsubsection{Evolving Model}
\label{sub:comp}

The evolving model allows CT AGN to evolve independently 
of the less obscured type 2 AGN.  Instead, $f_{CT}$ is assumed to be Eddington 
ratio dependent.  In order to fit the peak of the CXRB at $\sim$30 keV,  
the local CT AGN space density, and the $z$ $>$ 1 IR CT AGN space density, it is 
found that $\sim$86$\%$ of AGN with Eddington ratios $>$0.9 are CT,  
$\sim$60$\%$ of AGN with Eddington ratios $<$ 0.01 are CT, and $\sim$0$\%$ of AGN 
with intermediate Eddington ratios are CT.  

\begin{figure*}
\begin{center}
\includegraphics[angle=0,width=0.95\textwidth]{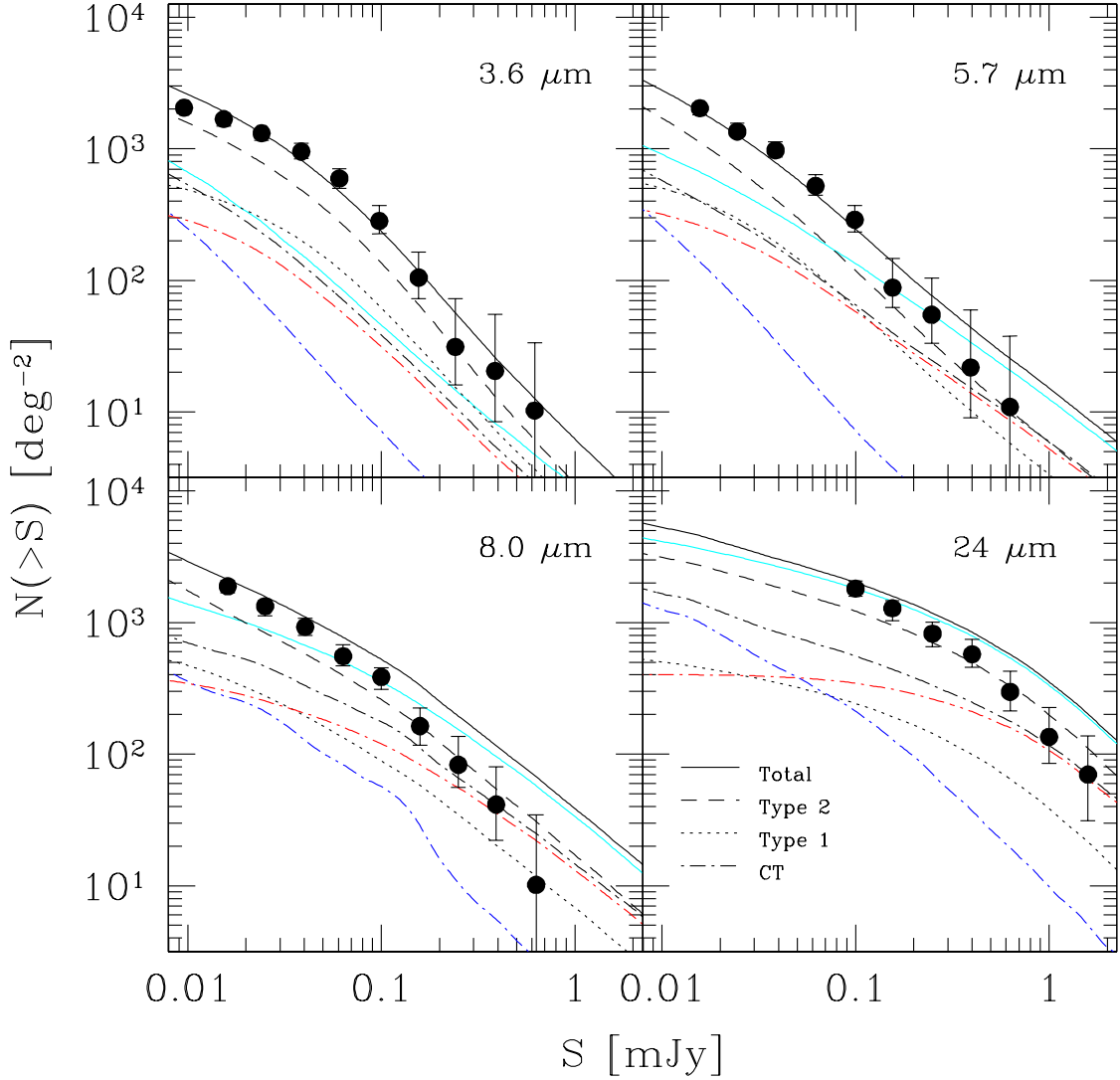}
\end{center}
\caption{Near and mid-IR number counts for AGN and hosts for the evolving model with $f_{2-8}$ $>$ 1 $\times$ 10$^{-16}$ erg s$^{-1}$ cm$^{-2}$.  As in Figure \ref{fig:ctsorig}, the cyan solid lines show the predicted number counts for AGN alone, the black solid lines show the total AGN and host number counts, the black dotted lines show the counts for type 1 AGN and hosts, and the black dashed lines show the counts for type 2 AGN and hosts.  The type 1 and type 2 AGN hosts are the same as in Figure \ref{fig:ctsorig}. The black dot-dashed lines show the total CT AGN and host number counts for the evolving model. The low Eddington ratio CT AGN hosts are shown by the blue dot-dashed lines and the high Eddington ratio CT AGN hosts are shown by the red dot-dashed lines.  The evolving model CT AGN host galaxies are summarized in Table \ref{work}.  The data shown is the same as in Figure \ref{fig:ctsorig}.}
\label{fig:ctscomp}
\end{figure*}

As AGN with weaker accretion rates are observed to have 
older stellar populations \citep{K03,K04,K07}, the low Eddington ratio AGN are assumed to 
have the same mass distribution as the type 1 and type 2 AGN but an older stellar 
population and very little star formation.  The age of the low Eddington ratio CT 
AGN host stellar populations are summarized in Table \ref{ages}.  If the low 
Eddington ratio CT AGN hosts have the same $E(B-V)$ as the type 1 and 
type 2 AGN hosts, the near and mid-IR number counts are over-predicted.  Increasing the 
dust in the low Eddington ratio CT AGN hosts to $E(B-V)$ $\approx$ 1.0 brings the 
model predictions into agreement with the faint end of the IR number counts, as shown 
by the blue dot-dashed lines in Figure \ref{fig:ctscomp}.  The larger amount of dust in the CT 
AGN hosts as compared to the type 2 AGN hosts makes sense as only galaxies containing a large 
amount of dust will be able to host an AGN with CT levels of obscuration. 

In the evolving model, the high Eddington ratio CT AGN hosts should be galaxies in a 
phase of rapid star formation and black hole growth.  Therefore, the high Eddington ratio CT 
AGN hosts should be LIRGs or ULIRGs.  Indeed we find that if the stellar populations of the 
high Eddington ratio CT AGN hosts have $L_{IR}$ = 10$^{12}$ L$_{\odot}$ from star 
formation, $E(B-V)$ $\approx$ 0.6, with a stellar population of age $\sim$0.3 Gyr which dominates 
$M_*$, in agreement with the average stellar population of ULIRGs which host an AGN 
\citep{RZ10}, the near and mid-IR number counts predictions are in agreement with deep 
observations.  The high Eddington ratio CT AGN and host number counts are shown as 
the red dot-dashed lines in Figure \ref{fig:ctscomp}.  The over-prediction at the bright end 
of the number counts is due to the incompleteness of the survey at bright fluxes.  As the GOODS 
fields cover only 0.1 deg$^2$ \citep{T06}, GOODS misses bright, rare objects.  If the stellar population is 
assumed to be younger than $\sim$0.3 Gyr, the number counts are over-predicted.  Similarly if 
the $E(B-V)$ $<$ 0.6 or if the average $M_*$ $>$ 10$^{11}$ M$_{\odot}$, the number counts are 
over-predicted.  Thus, the stellar population described here is the upper limit for how bright the 
high Eddington ratio CT AGN hosts can be without over-predicting the IR number counts. 

\begin{figure*}
\begin{center}
\includegraphics[angle=0, width=0.95\textwidth]{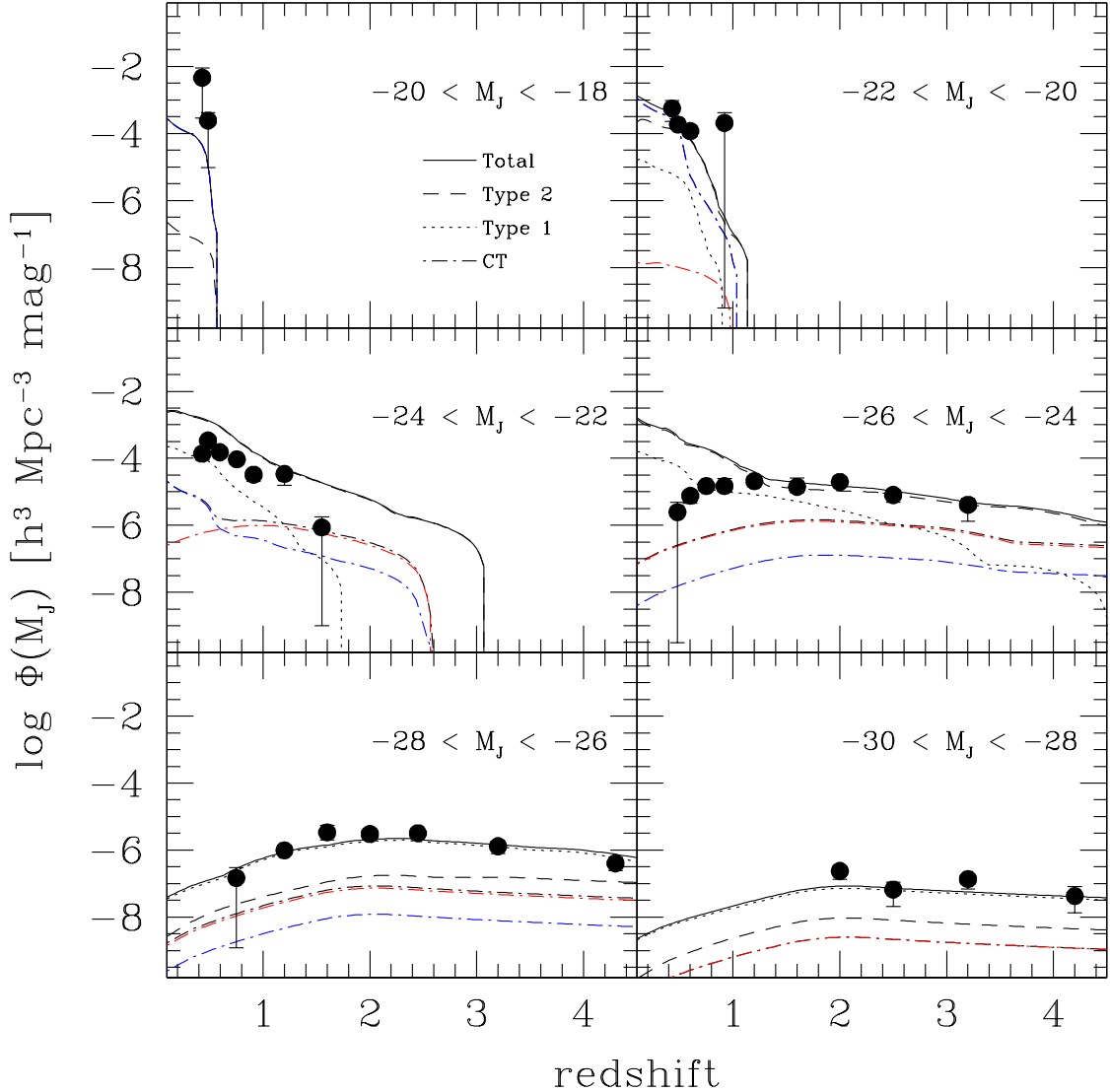}
\end{center}
\caption{J band space density for AGN and hosts for the evolving model.  The solid lines show the total AGN and host J band space density.  The dotted lines show the type 1 AGN and hosts space density and the dashed lines show the type 2 AGN and hosts space density.  The type 1 and type 2 AGN hosts have the same stellar populations as in Figure \ref{fig:jorig}.  The black dot-dashed lines show the total CT AGN and host J band space density for the evolving model. The low Eddington ratio CT AGN hosts, shown by the blue dot-dashed lines, and the high Eddington ratio CT AGN hosts, shown by the red dot-dashed lines, are described in Table \ref{work}.  Data points are the same as in Figure \ref{fig:jun}.}
\label{fig:jcomp}
\end{figure*}

\begin{figure*}
\begin{center}
\includegraphics[angle=0,width=0.95\textwidth]{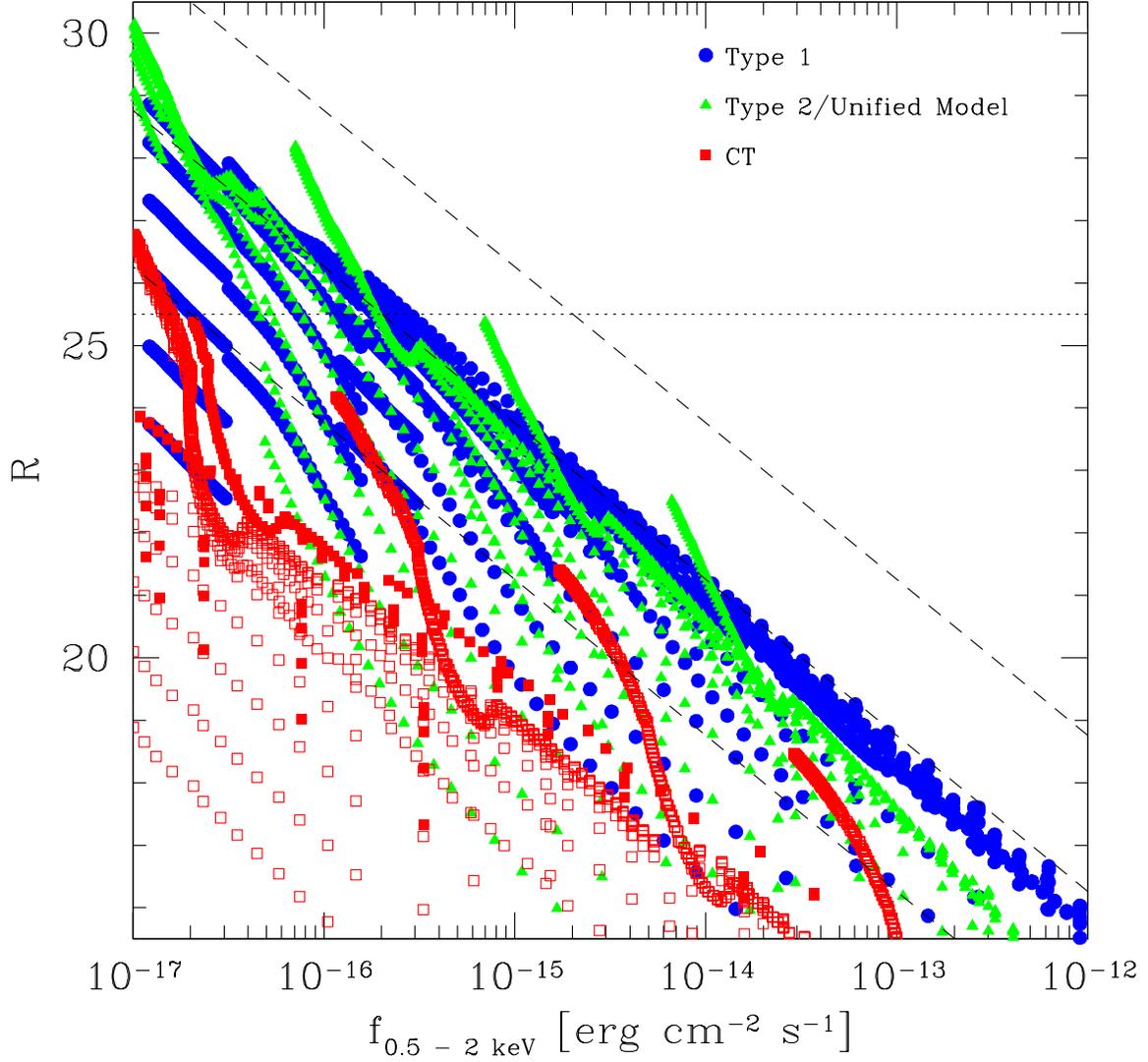}
\end{center}
\caption{Soft X-ray flux versus R band magnitude for AGN and hosts for the evolving model.  AGN and hosts are shown for $L_X$ $<$ 10$^{45}$ erg s$^{-1}$, $z$ $<$ 3, and $M_{*}$ = 10$^{9.5}$, 10$^{10}$, 10$^{10.5}$, 10$^{11}$, 10$^{11.5}$, and 10$^{12}$ M$_{\odot}$.  The blue circles show type 1 AGN and hosts and green triangles show type 2 AGN and hosts.  The type 1 and type 2 AGN hosts are the same as in Figure \ref{fig:fxtororig}.  The red squares show CT AGN and hosts.  The red filled squares show the low Eddington ratio CT AGN and hosts, while the open red squares show the high Eddington ratio CT AGN and hosts.  The evolving model CT AGN host galaxies are described in Table \ref{work}.  The horizontal dotted line again marks $R$ = 25.5, above which the source is considered an optically faint X-ray AGN.}
\label{fig:fxtorcomp}
\end{figure*}

The CT AGN host galaxies are also in good agreement with the other model 
constraints considered here.  The CT AGN host contribution to the J band 
space density is shown as the dot-dashed lines in Figure \ref{fig:jcomp}, where the 
red lines refer to the high Eddington ratio CT AGN and their hosts and the 
blue lines refer to the low Eddington ratio CT AGN and their hosts.  As 
expected the high Eddington ratio sources dominate the CT AGN and host 
contribution in the brighter magnitude bins while the low Eddington ratio sources 
dominate at the fainter magnitude bins.  The optical colors are also in agreement 
with observations.  The low Eddington ratio CT AGN hosts are on average 
redder and fainter than the type 1 and type 2 AGN hosts, while the high Eddington 
ratio CT AGN hosts are on average as bright or slightly brighter than 
the type 1 and type 2 AGN hosts but on average a little redder than the type 1 and 
type 2 AGN hosts.  As with the non-evolving model, most of the CT AGN hosts 
lie below the empirical average $\log(f_{0.5-2}/f_R)$ ratio, as expected by the high 
levels of soft X-ray absorption fundamental to CT AGN.  In Figure 
\ref{fig:fxtorcomp} the $f_{R}$ versus $f_{0.5-2}$ for CT AGN and their hosts are shown as the red squares 
with the low Eddington ratio objects shown as filled red squares and the high Eddington objects 
shown as open red squares.  The same result is found when the host galaxy IMF or metallicity is varied.

As the high Eddington ratio CT AGN are a small fraction of the overall 
AGN population, the evolving model CT AGN hosts are not affected by the consideration of the 
enhanced star formation sources.  Thus the CT AGN hosts of the evolving model 
are in agreement with the AGN evolution scenario where major mergers trigger 
nuclear starbursts and highly obscured AGN activity.

\subsection{Evolution of $f_{2}$}
\label{sub:evf2}

The evolution of $f_2$, the type 2 AGN fraction, is important for understanding the AGN 
life cycle and how AGN and their host galaxies interact \citep[e.g.,][]{B06a}.  In the 
unified model, $f_2$ is the covering factor 
of the dusty torus, and thus the evolution of $f_2$ shows a fundamental evolution of the 
torus parameters \citep{B06}.  Several studies suggest that $f_2$ evolves with $L_X$ and 
possibly also with redshift \citep[e.g.,][]{U03, S05, B06a, H08, F09, W09}.  However, other studies 
find that it is not necessary for $f_2$ to evolve with $L_X$ \citep[e.g.,][]{LE10} nor 
with redshift \citep[e.g.,][]{GF03, TU05} in order to explain observations.
Indeed, we find that the number counts and J band space density 
can be fit with a constant $f_2$.  For $z$ $<$ 0.4, the type 1 AGN and host B band luminosity 
function can also be satisfactorily fit with a constant $f_2$.  However, fitting the type 1 AGN 
and host B band luminosity function at $z$ $\gtrsim$ 1 requires that $f_2$ is not constant.  
If $f_2$ is constant, the best fit $\chi_{red}^2$ $>$ 7.0 and the type 1 AGN and host B band luminosity function 
predicts a considerably larger population of B band magnitude, $m_B$ $\sim$ 23, 1 $<$ $z$ $<$ 2 type 1 AGN than 
observed.  Thus, it is found that in order to fit the type 1 AGN and host B band luminosity function at $z$ $>$ 
1, $f_2$ must evolve with $L_X$.  

\begin{figure*}
\begin{center}
\includegraphics[angle=0,width=0.95\textwidth]{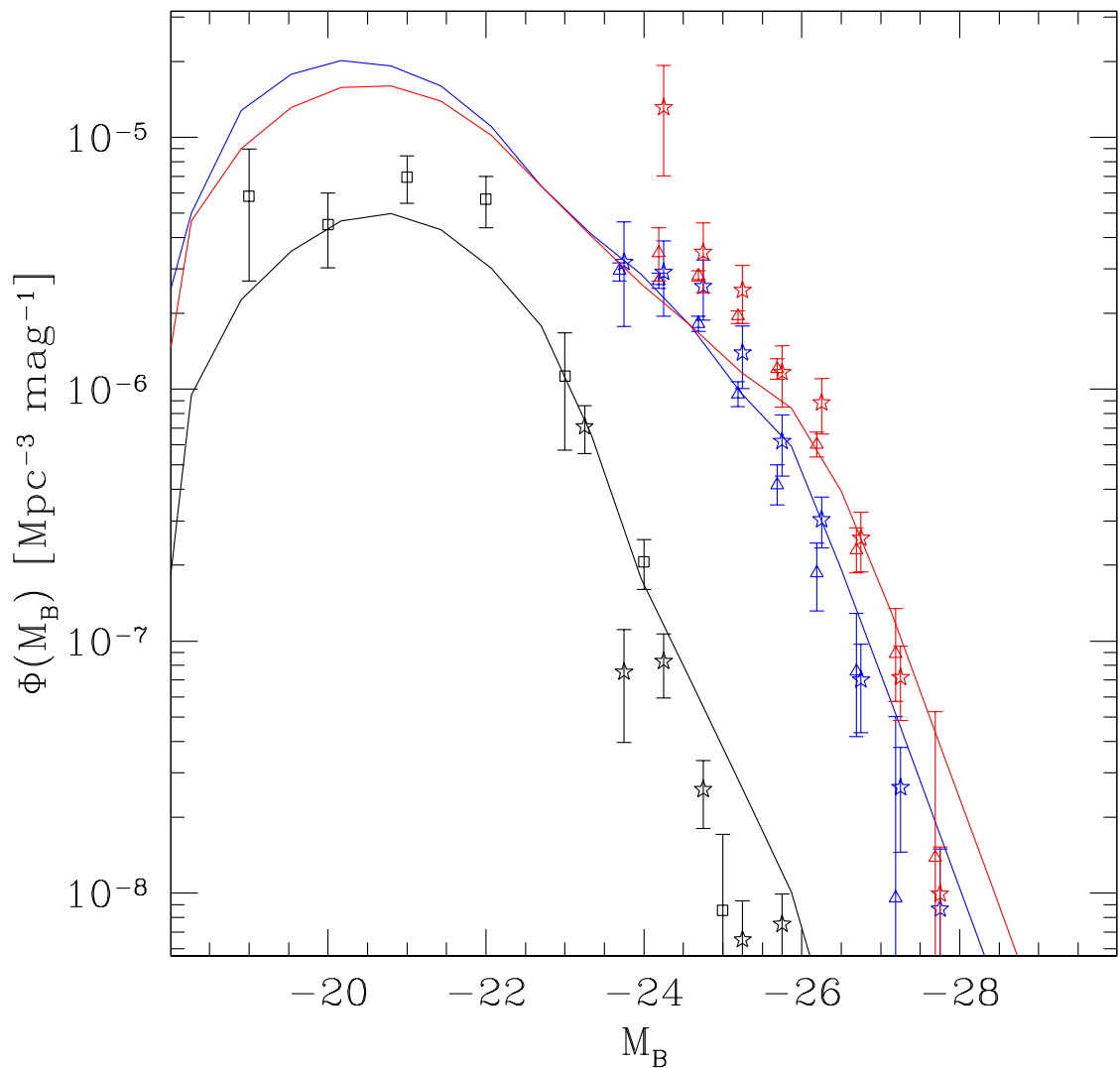}
\end{center}
\caption{Type 1 AGN and host B band luminosity function when $f_2$ is assumed to not evolve with $z$.  Colors and data points are the same as in Figure \ref{fig:bband}.  The $z$ $<$ 1 type 1 AGN hosts have $E(B-V)$ $\approx$ 0.25 and at higher redshift the type 1 AGN hosts have $E(B-V)$ $\approx$ 0.4, in contrast to the case where $f_2$ does evolve with $z$ and type 1 AGN hosts have a lower $E(B-V)$ at higher $z$ than locally.}
\label{fig:bf2Lx}
\end{figure*}

The argument for evolution of $f_2$ with redshift is less conclusive. 
Figure \ref{fig:bf2Lx} shows the type 1 AGN and host B band luminosity function 
assuming that $f_2$ does not evolve with redshift, which gives $\chi_{red}^2$ = 6.5.  
When $f_2$ does evolve with redshift, the best fit $\chi_{red}^2$ = 4.3.  While the $\chi_{red}^2$ 
for both the evolving and the non-evolving $f_2$ models do not represent formal good fits to the 
observed data points, the model in which $f_2$ does evolve with redshift provides a better fit 
to the data. In the scenario where $f_2$ does not evolve with redshift, it is necessary 
that at higher redshift the type 1 AGN host galaxies be dustier than locally.  This 
is in contrast to the scenario where $f_2$ does evolve with redshift and type 1 AGN hosts are 
less dusty at higher redshift compared to the local population.  In order to not over-predict the IR number counts, 
the type 2 AGN hosts must also be dustier at higher redshift than locally.  Assuming that $f_2$ 
does not evolve with redshift has minimal affect on the J band space density, optical colors, and 
$\log(f_{0.5-2}/f_R)$ ratio.  

A comparison of Figure \ref{fig:bband} and Figure \ref{fig:bf2Lx}, suggests that $f_2$ 
not only evolves with $L_X$, but also with redshift.  Also, the model where $f_2$ evolves 
with both $L_X$ and redshift provides a better fit to the type 1 AGN and host B band luminosity 
function has measured by $\chi_{red}^2$.  As neither the evolving nor non-evolving $f_2$ models 
provide a formally good fit to the observed luminosity function, the possibility that $f_2$ does 
not evolve with redshift cannot be conclusively ruled out; however, the evolution of the B band 
type 1 AGN and host luminosity function suggests that $f_2$ does evolve with redshift.  In order 
to fit the type 1 AGN and host B band luminosity function $f_2$ must evolve with $L_X$.

\subsection{Summary of Results}
\label{sub:sumres}

Using a variety of optical and near and mid-IR data, we have constrained the stellar populations 
of AGN host galaxies.  Table \ref{work} summarizes the average AGN host galaxy for different AGN 
spectral types.  It is found that at $z$ $<$ 1 type 1 and type 2 AGN hosts are similar, but at 
$z$ $>$ 1 type 1 AGN hosts are less dusty than type 2 AGN hosts.  The majority of AGN have an 
average SFR $\lesssim$ 2 M$_{\odot}$ yr$^{-1}$, however there is evidence of a population of 
enhanced star formation sources which account for $\gtrsim$ 5--15$\%$ of the AGN population 
and has SFR $\approx$ 20 M$_{\odot}$ yr$^{-1}$.  Also, it is found that if CT AGN evolve like 
type 2 AGN, then CT AGN hosts are similar to type 2 AGN hosts and if CT levels of obscuration 
are indicative of specific evolutionary stages in the AGN life cycle, then CT AGN hosts are also 
in specific evolutionary stages.  Furthermore, it is shown that $f_2$ evolves with $L_X$ and $f_2$ 
is likely to evolve with redshift.

\section{Discussion}
\label{sect:disc}

\subsection{Hosts of CT AGN}
\label{sub:CT}

For both the evolving and  
non-evolving models, the CT AGN host stellar populations suffer 
from at least as much dust extinction as the type 2 AGN host stellar populations.  
Observations suggest that at least some of the observed extinction 
of heavily obscured AGN may be due to extended dust structures or 
molecular clouds within the host galaxy \citep{B07, P08, MS10}, so 
it is expected that the stellar populations of CT AGN 
will be enshrouded in dust.  It is within the limits imposed 
by the model constraints for the high Eddington ratio CT 
AGN hosts to have an average $E(B-V)$ which is in agreement with ULIRGs 
hosting AGN \citep{RZ10}.

The AGN evolution scheme, in which mergers trigger large nuclear 
starbursts and AGN activity, claims that the AGN activity will initially 
be very highly obscured while the black hole grows very rapidly.  As the 
black hole grows, the radiation pressure on surrounding dusty gas will increase until the AGN 
feedback blows out the obscuring material and halts the star formation 
in the host nuclear region \citep{S88, P04, Ri09, H06}.  In this scheme 
CT AGN should have young stellar populations and possibly 
high levels of on going star formation.  The non-evolving model places a 
lower limit on the stellar population age of 1 Gyr.  However, when the 
enhanced star formation sources are included, the non-evolving model CT 
AGN hosts have similar stellar ages to the type 2 AGN hosts, in 
contrast to the expectations of the AGN evolution scheme.  For the high 
Eddington ratio CT AGN in the evolving model, the lower 
limit on the stellar population age is 0.3 Gyr, the average 
stellar age of ULIRGs hosting an AGN \citep{RZ10}.  The low Eddington 
ratio CT AGN hosts of the evolving model can be of similar 
age as the type 1 and type 2 AGN, but the near and mid-IR predicted number 
counts are in better agreement with the observations if these AGN hosts have 
slightly older stellar populations than the average type 1 and 
type 2 AGN hosts. It has 
been demonstrated that galaxies hosting AGN with lower [OIII] 
luminosities have larger values for $D_n$(4000 \AA) compared to 
galaxies hosting AGN with high [OIII] luminosities \citep{K03,K04,K07}.  
Thus it is expected that AGN with lower Eddington ratios are 
in hosts with older stellar populations, in agreement with 
the findings of this study.    

The stellar populations of CT AGN hosts for the evolving model and 
the non-evolving model are in agreement with the expectations from the AGN 
evolution scheme.  The non-evolving model finds that the average CT 
AGN host has recently ($\gtrsim$1 Gyr ago) undergone a large burst of star 
formation, but that current star formation rates are more modest.  The 
evolving model finds that high Eddington ratio CT AGN hosts have recently 
($\gtrsim$0.3 Gyr ago) undergone a large burst of star formation and that 
current star formation rates may also be elevated.  If the 
enhanced star formation sources are included, the non-evolving model AGN hosts 
have stellar populations of similar age as the type 2 AGN, and thus are in 
better agreement with the orientation based unified model than with the AGN 
evolution scheme.  The evolving model CT AGN hosts are consistent with 
the AGN evolution scheme regardless of the inclusion of the  
enhanced star formation sources.

In summary, the non-evolving model CT AGN hosts are a simple extension of the type 
2 AGN host population while the evolving model CT AGN are consistent with the paradigm 
where major mergers cause both intense starbursts and AGN activity.  In the evolving model, 
low Eddington ratio CT AGN hosts will appear as galaxies with old and dusty stellar populations 
while the high Eddington ratio CT AGN hosts will appear as IR bright starburst galaxies.

\subsection{Enhanced Star Formation in AGN Hosts}
\label{sub:ensf} 

It is found that $\sim$5-15$\%$ of AGN hosts can be enhanced star formation sources 
with an average SFR $\approx$ 20 M$_{\odot}$ yr$^{-1}$, a factor of 10 higher 
than the majority of AGN hosts.  This is expected to be a lower limit of the fraction of 
enhanced star formation sources as some sources may 
have such highly embedded star formation that the majority of the reprocessed emission due 
to star formation is at wavelengths longer than 24 $\mu$m.  Observations at longer wavelengths, 
such as in the far-IR with the {\em Herschel} 
or the millimeter/sub-mm regime with the Atacama Large Millimeter/submillimeter Array (ALMA), are 
necessary for uncovering the evolution of star formation rates in AGN hosts \citep{DB11}.  Even deep radio observations 
are a useful tool in determining the highly embedded star formation rates of AGN hosts \citep{Bal09}.  
Indeed, by stacking sub-mm observations 
of X-ray selected AGN, \citet{L10} found that the average AGN host SFR $\approx$ 30 M$_{\odot}$ 
yr$^{-1}$.   
Furthermore, \citet{L10} found  that the AGN host SFR evolves strongly with redshift but 
with evidence of luminosity dependent evolution only for the highest luminosity AGN.  In the scenario 
where AGN host SFRs are not luminosity dependent, it is found that more than half of AGN hosts 
can be enhanced star formation sources.


The fact that the mid-IR AGN and host number counts are over-predicted by an average SFR $>$ 2 
M$_{\odot}$ yr$^{-1}$, despite observational evidence that there is a population of AGN with an average 
SFR an order of magnitude higher than this upper limit, suggests that there are two 
populations of AGN.  The majority of AGN hosts, at least at $z$ $<$ 1, have SFRs 
similar to local spiral galaxies and $\gtrsim$5-15$\%$ of AGN hosts have markedly higher 
SFRs.  The existence of these two populations of AGN hosts can be interpreted 
in two complimentary ways.  The first interpretation is that the AGN hosted by galaxies with enhanced SFRs 
are being fueled by different mechanisms than the lower star formation objects.  
At $z$ $<$ 1, the lower star formation objects are likely dying quasars which were triggered by major 
mergers while the enhanced star formation objects are likely Seyferts which are both fueled and obscured by 
circumnuclear starburst disks, as 
investigated by \citet{B08}.  The other interpretation is that all AGN are fueled by processes related to nuclear starbursts,
whether those starbursts are triggered by mergers or through secular processes, 
and on average, the nuclear starburst phase overlaps with the active AGN phase for 
$\gtrsim$5-15$\%$ of the AGN lifetime.  Further exploration of these two populations 
of AGN hosts is necessary to determine the processes which trigger and fuel AGN activity.

\subsection{Methods for Finding CT AGN}
\label{sub:findCT}


It is well documented that the integrated emission of AGN observed in deep X-ray surveys 
is insufficient to account for the intensity of the CXRB at $\sim$30 
keV and that the shape of the CXRB necessitates that the missing population 
of AGN be highly obscured \citep[e.g.,][]{B06a, DB09, T09b}. Given the uncertainties 
of the normalization of the CXRB and the AGN hard X-ray luminosity functions, 
different models predict vastly different numbers of missing CT AGN \citep[e.g.,][]{G07,DB09,T09b}.  
As these highly obscured AGN are missed in deep X-ray surveys \citep{H08}, 
it is important to consider other methods to identify the elusive CT AGN population.

One possibility is that the majority of X-ray bright optically inactive galaxies (XBONGs) 
host CT AGN \citep{Fi08,Fi09,T09,R10}.  XBONGs are X-ray sources found in deep surveys 
which have no optical counterparts with $R$ $\lesssim$ 25.5 and make up a 
substantial portion of deep survey X-ray sources \citep{A10}.  It is thought that these sources 
are either heavily obscured AGN and/or high redshift quasars 
\citep{A01,M05,R10}.   Considering the sources above the dotted horizontal 
line in Figures \ref{fig:fxtororig} 
and \ref{fig:fxtorcomp} suggests that a small fraction of 
optically faint X-ray sources may be CT AGN, but the majority 
of this population is $z$ $\gtrsim$ 1 type 2 AGN.  Thus it is unlikely that XBONGs 
host the majority of the missing CT AGN population.

Another method used to identify CT AGN candidates is to search for infrared 
bright sources which have an infrared excess, usually defined by $f_{24}/f_R$ 
$\gtrsim$ 1000, where $f_{24}$ is the 24 $\mu$m flux \citep[e.g.][]{P06, A08, 
D08, Fi08, Fi09, T09}.  Some concern has been raised that this method will also 
pick out lower redshift type 2 AGN masquerading as high redshift CT AGN \citep{G10}.  
We therefore investigate the population of AGN selected by the infrared excess criteria.

\citet{Fi08} suggest that using the criteria $f_{24}/f_R$ $\gtrsim$ 1000 and $R$ - $K$ $>$ 
4.5 selects a distinct class of sources, the majority of which are CT AGN.  However, according 
to both the evolving and non-evolving model CT AGN and host SEDs developed here, the \citet{Fi08} 
criteria selects low to moderately X-ray bright AGN at moderate to high redshift.  The vast majority 
of the selected AGN are obscured, but a significant fraction are still Compton thin.  Similarly, 
\citet{D08} find that the \citet{Fi08} criteria is likely to select dusty star forming templates 
and low X-ray flux AGN.  However, \citet{D08} find no evidence that the selected AGN are CT.

\begin{figure*}
\begin{center}
\includegraphics[angle=0,width=0.95\textwidth]{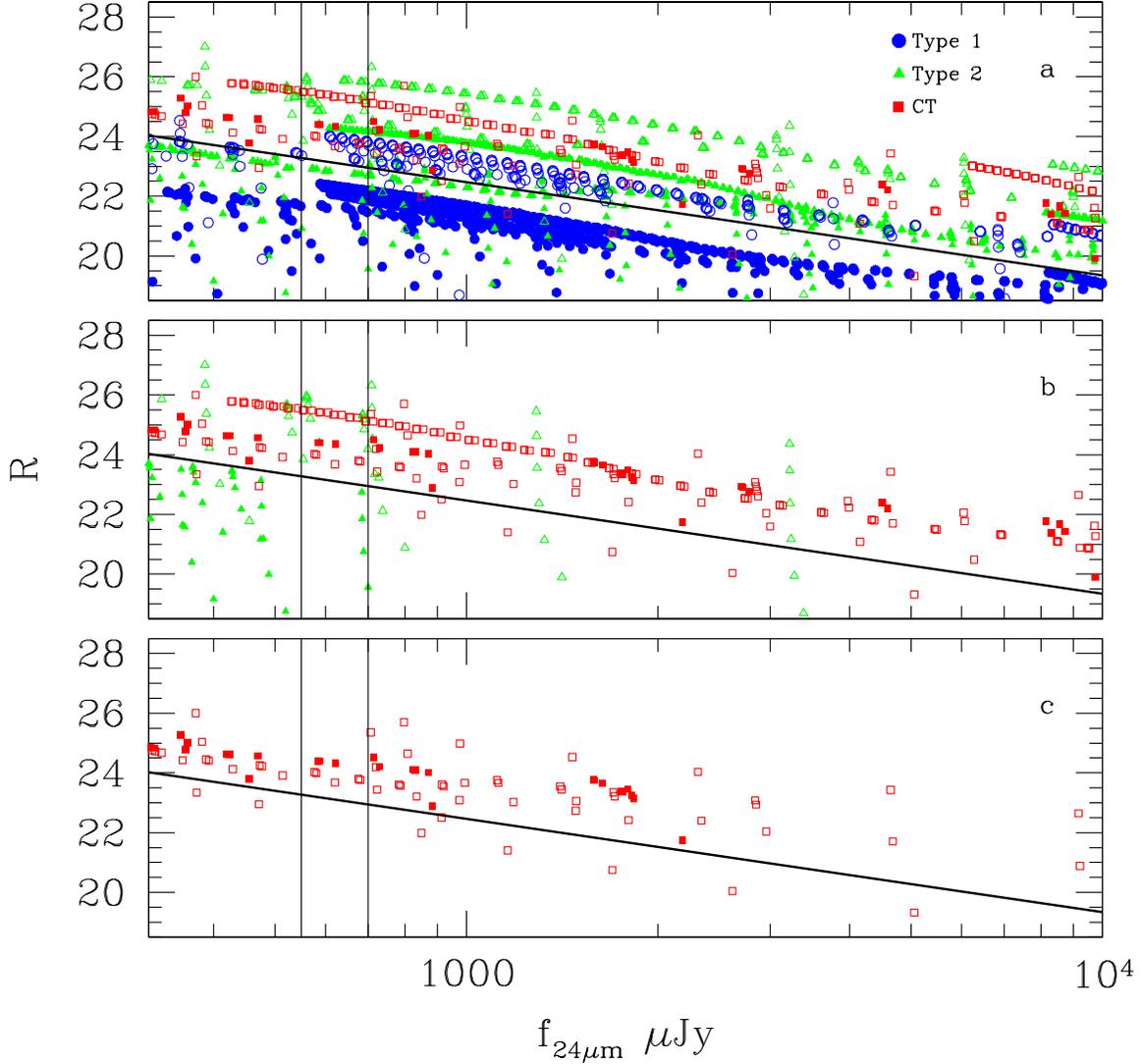}
\end{center}
\caption{$R$ versus f$_{24}$ for evolving model AGN with $L_X$ = 10$^{42}$, 10$^{43.5}$, 10$^{45}$ and 10$^{46.5}$ erg s$^{-1}$, $z$ $<$ 5, and $M_{*}$ = 10$^{10}$, 10$^{11}$, and 10$^{12}$ M$_{\odot}$ for various soft X-ray flux ranges.  Panel a shows all AGN regardless of $f_{0.5-2}$, panel b shows all AGN with $f_{0.5-2}$ $<$ 10$^{-15}$ erg s$^{-1}$ cm$^{-2}$, and panel c shows all AGN with $f_{0.5-2}$ $<$ 10$^{-16}$ erg s$^{-1}$ cm$^{-2}$.  The blue circles show the type 1 AGN and the green triangles show the type 2 AGN.  The open points mark the enhanced star formation objects.  The red filled squares show the low Eddington ratio CT AGN and the open red squares show the high Eddington ratio CT AGN. The thick black line shows where $f_{24}$/$f_{R}$ = 1000, therefore the area of interest is above the thick line.  The vertical lines mark $f_{24}$ = 550 $\mu$Jy and $f_{24}$ = 700 $\mu$Jy.}
\label{fig:f24tor}
\end{figure*}

\begin{figure*}
\begin{center}
\includegraphics[angle=0,width=0.95\textwidth]{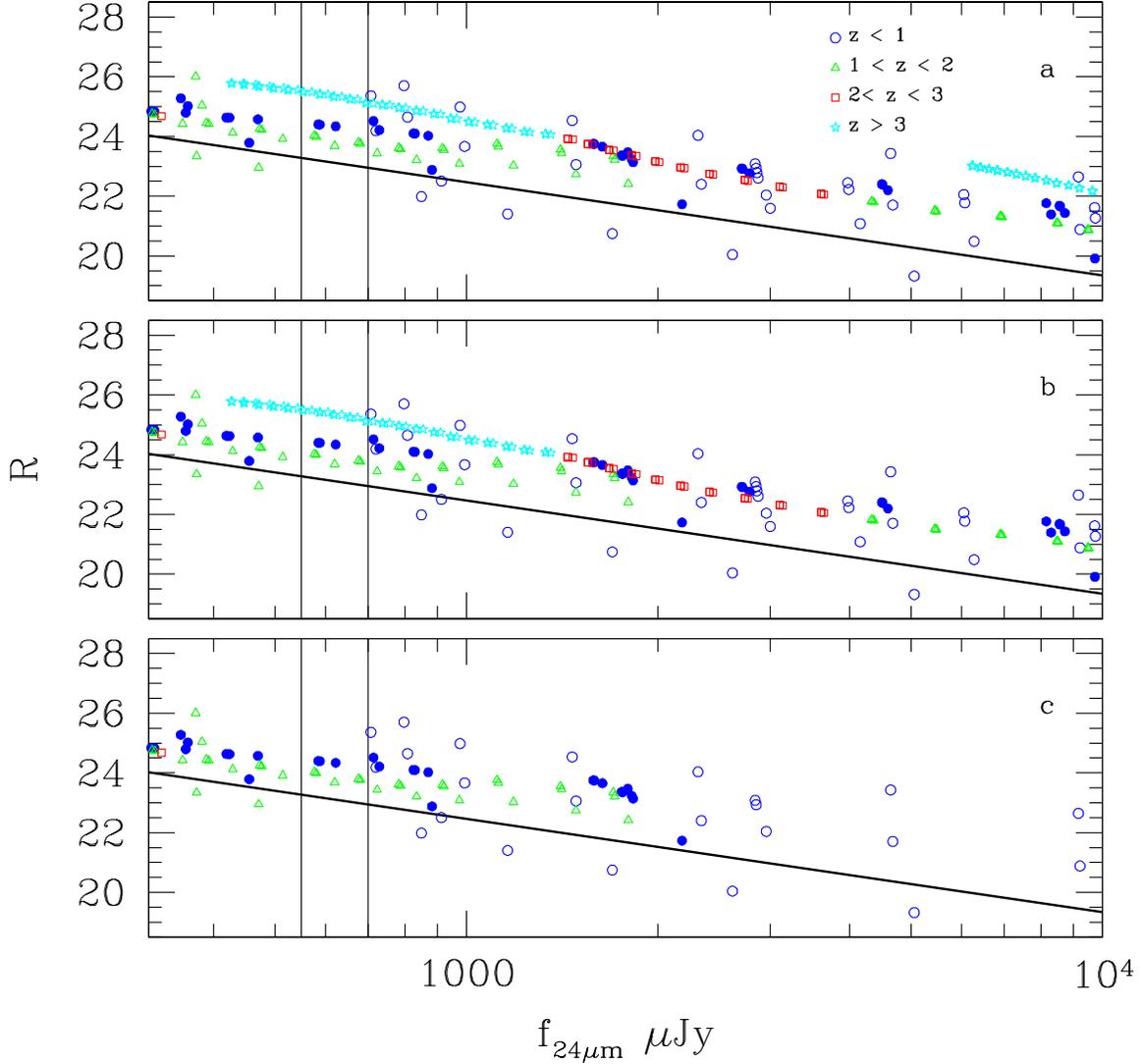}
\end{center}
\caption{Redshift distribution of $R$ versus $f_{24}$ for CT AGN with $L_X$ = 10$^{42}$, 10$^{43.5}$, 10$^{45}$ and 10$^{46.5}$ erg s$^{-1}$ and $M_{*}$ = 10$^{10}$, 10$^{11}$, and 10$^{12}$ M$_{\odot}$ using the evolving model for various soft X-ray flux ranges.  Panel a shows all CT AGN regardless of $f_{0.5-2}$, panel b shows the CT AGN with $f_{0.5-2}$ $<$ 10$^{-15}$ erg s$^{-1}$ cm$^{-2}$, and panel c shows the CT AGN with $f_{0.5-2}$ $<$ 10$^{-16}$ erg s$^{-1}$ cm$^{-2}$.  Point styles designate AGN in different redshift ranges.  Blue circles show AGN with $z$ $<$ 1, green triangles show AGN with 1 $<$ $z$ $<$ 2, red squares show AGN with 2 $<$ $z$ $<$ 3, and cyan stars show AGN with $z$ $>$ 3.  The filled points show the low Eddington ratio CT AGN and the open points show the high Eddington ratio CT AGN. The thick black line shows where $f_{24}$/$f_{R}$ = 1000, therefore the area of interest is above the thick line.  The vertical lines mark f$_{24}$ = 550 $\mu$Jy and f$_{24}$ = 700 $\mu$Jy.}
\label{fig:f24torbyz}
\end{figure*}

A more effective method of identifying CT AGN candidates is based on the selection 
criteria of \citet{P08}: $f_{24}/f_R$ $\gtrsim$ 1000 and $f_{24}$ $>$ 1.0 mJy.  Furthermore, it 
appears that samples with a lower $f_{24}$ limit, such as  $f_{24}$ $>$ 700 $\mu$Jy or even 
$f_{24}$ $>$ 550 $\mu$Jy, also contain a large fraction of highly obscured AGN \citep[][and 
references therein]{D10}.  Fainter samples of infrared bright galaxies 
are found to be predominately powered by star formation rather than by AGN  
\citep[e.g.,][]{Pope08}.  Figure \ref{fig:f24tor}a shows that 
representatives of all spectral types of AGN can be found with $f_{24}/f_R$ 
$\gtrsim$ 1000 and $f_{24}$ $>$ 550 $\mu$Jy.  These sources will be located at all redshifts.  
However, the majority of type 1 and type 2 AGN are bright in the soft X-ray.  
For both the evolving and non-evolving models, 
the vast majority of infrared excess AGN with $f_{24}$ $>$ 550 $\mu$Jy 
and f$_{0.5-2}$ $<$ 10$^{-15}$ erg cm$^{-2}$ s$^{-1}$ are CT AGN, as shown in 
Figure \ref{fig:f24tor}b.  Figure \ref{fig:f24torbyz}b, which 
shows the redshift distribution of the CT AGN with f$_{0.5-2}$ $<$ 10$^{-15}$ erg cm$^{-2}$ s$^{-1}$, 
illustrates that these X-ray faint CT AGN are located at all redshifts.
For the evolving model the vast majority of 
these CT AGN are high Eddington ratio sources.  
This criteria will select a small number of type 2 AGN 
with enhanced star formation, but the majority of sources selected in this manner are indeed CT.    

If an infrared excess AGN has $f_{0.5-2}$ $\lesssim$ 10$^{-16}$ erg 
cm$^{-2}$ s$^{-1}$ and $f_{24}$ $>$ 550 $\mu$Jy, than the AGN is CT 
and located at $z$ $\lesssim$ 2.  This is true for both the evolving 
and non-evolving models. An interesting consequence of this is that a size-able 
population of $z$ $>$ 2 CT AGN should have 10$^{-15}$ $>$ $f_{0.5-2}$ 
$>$ 10$^{-16}$ erg s$^{-1}$ cm$^{-2}$.  According to the evolving model, nearly 
all of the $f_{0.5-2}$ $\lesssim$ 10$^{-16}$ erg cm$^{-2}$ s$^{-1}$ sources are 
high Eddington ratio sources and according to the non-evolving model 
these low X-ray flux sources are nearly all enhanced star formation sources.    
It is therefore expected that X-ray stacking of IR bright sources with 
$f_{24}/f_R$ $\gtrsim$ 1000 and low soft X-ray flux will yield a large fraction 
of the $z$ $\lesssim$ 2 high luminosity CT AGN population.  
This shows that combining observations in multiple spectral regions, such as mid-IR, 
optical, and X-ray observations, is the 
most efficient way of identifying CT AGN candidates.

Identifying and characterizing the elusive CT AGN population is a necessary part 
of understanding the history of accretion and galaxy evolution.  However, due to 
the observational challenges of studying highly obscured sources, multi-wavelength 
investigations are necessary to identify and understand the nature of CT AGN.

\subsection{Implications for the Unified Model and AGN Fueling Mechanisms}
\label{sub:ave}

Several recent studies suggest that there are two distinct processes which 
lead to AGN activity, secular evolution and merger events \citep[e.g.,][]{B06a,H08,HH09,L10}.  
In the latter paradigm, quasar activity is activated by galaxy mergers which cause gas 
and dust to be funneled into the nuclear region \citep[e.g.,][]{S88,F99,P04,H06}, 
whereas moderate luminosity AGN are fueled by gravitational instabilities internal 
to the host galaxy or through minor interactions \citep[e.g.,][]{Cr03,KK04,P07,S08}.  It 
is probable that if different forms of AGN activity are caused by contrasting fueling mechanisms, 
the relationship between various AGN spectral types may be different for the high and 
moderate luminosity populations.  

This study finds that the average type 1 and type 2 AGN hosts have similar stellar populations 
and similar levels of dust attenuation at $z$ $<$ 1.  At $z$ $>$ 1 type 1 and type 2 
AGN hosts appear to be fundamentally different.  Even though the high redshift type 1 and type 
2 hosts have similar stellar populations, the type 1 AGN and host B band luminosity 
function requires that type 1 AGN hosts have significantly lower levels of dust extinction 
than what is required for type 2 AGN hosts to be in agreement with the observed near and mid-IR number 
counts.  At $z$ $>$ 1, type 1 AGN hosts are intrinsically less dusty than type 2 AGN hosts.  
This suggests that the orientation based unified model works well for describing the local 
Seyfert population, but may not be appropriate for the high redshift, high luminosity  
quasar population.  This conclusion is consistent with the findings of several other recent 
studies \citep[e.g.,][]{B06,HH09,DB10,L10}.  For galaxies which are evolving secularly, the 
unified model appears to be an apt description.  However, the violent growth experienced by 
black holes and their host bulges during major merger events does not appear to fit into the 
orientation based unified scheme.

It is likely that different AGN fueling mechanisms will result in different relationships 
between AGN spectral types.  In the merger scenario, young AGN are highly obscured and old 
quasars are unobscured \citep[e.g.,][]{S88, F99,P04,H06}.  Different spectral types of AGN 
fueled by secular processes related to nuclear starbursts are expected to be in agreement with the orientation 
based unified model due to the disky nature of the nuclear starburst which is likely to both fuel and 
obscure the AGN \citep[e.g.,][]{B08}.  In the evolving model, it is found that mergers can play a 
strong role in fueling high $L_X$, $z$ $>$ 1 quasars.  However, in the non-evolving 
model, when the enhanced star formation sources are considered, it appears that the dominant 
fueling mechanism is not mergers.  Instead, the difference between spectral types may only be 
that some galaxies have less dust and gas than others at $z$ $>$ 1.  In the non-evolving 
$f_{CT}$ model, it appears that secular processes are the dominate AGN fueling mechanism at all redshift.  
For both the evolving and non-evolving model, at $z$ $<$ 1 the processes which lead to AGN activity are most likely secular.  However, the 
findings of this study suggest that the dominant AGN fueling process changes at $z$ $\sim$ 1, since  
at $z$ $<$ 1 the average type 1 and type 2 AGN hosts are more similar than at higher redshift.  In order 
to understand the mechanisms which trigger and fuel AGN, it is important for future studies to pay 
careful attention to which fueling mechanisms are dominant in different subsets of the AGN population.

Thus, it has been shown that the relationship between different AGN spectral types is a helpful tool for understanding 
the dominant processes responsible for triggering and fueling AGN activity.  It is found that at 
$z$ $<$ 1 the orientation based unified model holds, suggesting that the 
dominant AGN fueling mechanisms at $z$ $<$ 1 are secular processes.  At $z$ $>$ 1, the orientation 
based unified model does not seem to hold.  The evolving CT AGN model suggests that at $z$ $>$ 1 
mergers are an important AGN fueling mechanism; however, the non-evolving CT AGN model suggests that 
even at $z$ $>$ 1 the dominant AGN fueling mechanisms are secular processes and that mergers play only 
a minor role in fueling AGN at all redshifts.


\subsection{$L_X$ and redshift Evolution of $f_2$}
\label{sub:evolf2}

Understanding the evolution of $f_2$ is an important step in understanding the nature of 
AGN obscuration and the interplay between an AGN and it's host galaxy.  If $f_2$ evolves 
with AGN $L_X$, this suggests that the obscuring material is close enough to the central 
engine that dust sublimation \citep{Law91} and/or radiation pressure 
fed winds \citep{KK94} affect portions of the obscuring material.  
If $f_2$ evolves with redshift this would indicate that the evolution of the obscuring 
material is somehow connected to the evolution of the host galaxy \citep{B06}.  While most 
model constraints used in this study are not very sensitive to the evolution of $f_2$, the 
type 1 AGN and host B band luminosity function does prove to be a valuable test of the 
evolution of $f_2$.  This study finds that $f_2$ must evolve with AGN $L_X$ and that 
$f_2$ probably evolves with redshift.

If $f_2$ does not evolve with $L_X$, the predicted 1 $<$ $z$ $<$ 2 type 1 AGN and host B 
band luminosity functions are not in agreement with observations.  Using the host galaxy 
stellar population age and $E(B-V)$ as free parameters does not allow for an appropriate 
fit to the data.  The over-prediction of $m_B$ $\sim$ 23 sources at 1 $<$ $z$ $<$ 2 is 
not due to observational bias as $m_B$ $\sim$ 23 is significantly brighter than the depth 
of B band coverage accessible to surveys \citep[e.g., B band 5$\sigma$ level for COSMOS 
is $m_B$ = 26.7;][]{Scoville07}.  \citet{LE10} argue that $f_2$ does not evolve with $L_X$ 
and that apparent evolution of $f_2$ with $L_X$ is due to X-ray observational biases.  
However, the $z$ $>$ 1 type 1 AGN and host B band luminosity function data used here is 
based on optical selection criteria \citep{HS90,C04}.  The \citet{DC96} type 1 AGN and 
host B band luminosity function does use a sample defined through X-ray selection, and 
the $z$ $<$ 0.4 predicted type 1 AGN and host B band luminosity function is in reasonable 
agreement with observation even if $f_2$ does not evolve with $L_X$.  This suggests that 
X-ray observational biases do not create an artificial evolution of $f_2$ with $L_X$.  
Therefore, at least for $z$ $>$ 1, $f_2$ must evolve with $L_X$. 

While this study cannot rule out the possibility that $f_2$ is constant with 
redshift, our findings suggest that $f_2$ does evolve mildly with redshift.  While some 
studies suggest that $f_2$ must evolve with redshift \citep[e.g.][]
{LF05,B06a,B06,TU06}, the evidence is still tentative.  If AGN are fueled 
by different mechanisms, one might assume that the type 1/type 2 ratio will be 
different for the different mechanisms.  Therefore, if different mechanisms 
dominate during different epochs, the type 1/type 2 ratio should also include 
some redshift evolution.  It is possible that the slower secular evolution 
processes of Seyfert galaxies do not require evolution of $f_2$ with redshift or 
that the evolution is very mild, since this form of galaxy evolution 
is governed by stochastic processes.  However, in the quasar regime, where AGN 
fueling is initiated by major mergers, a stronger evolution of $f_2$ with redshift 
may be necessary.  Indeed, the type 1 AGN and host B band luminosity function which 
is least well fit by a non-evolving $f_2$ is the highest redshift bin considered, which 
is the redshift bin closest to the peak of quasar activity.  While this study does not 
conclusively show that $f_2$ evolves with redshift, the findings presented here suggest 
that $f_2$ does exhibit some evolution with redshift.  

It is found that $f_2$ evolves with $L_X$ and $f_2$ is likely to evolve with redshift.  This 
suggests that the obscuring medium is close enough to the central engine to be affected 
by dust sublimation \citep{Law91} and/or radiation pressure fueled winds \citep{KK94}.  Also, the 
type 1 AGN and host B band luminosity function is an excellent X-ray independent tool to test the 
fraction of type 1 AGN predicted by X-ray observations.

\section{Summary}
\label{sect:sum}
By applying observational constraints at optical through mid-IR wavelengths to 
AGN host stellar population models, the average stellar properties of AGN hosts have been constrained 
and hard X-ray through mid-IR SEDs have been developed which include emission from the AGN and the host 
galaxy stellar population and dust enshrouded star formation\footnote{AGN and host SEDs are available by contacting the authors.}.
The findings of this study are summarized as follows.

\textbullet Type 1 and type 2 AGN hosts have similar stellar populations at $z$ $<$ 1.

\textbullet At $z$ $>$ 1, type 1 and type 2 AGN hosts have stellar populations of similar 
age, but type 1 AGN hosts are intrinsically less dusty than type 2 AGN hosts.   

\textbullet The orientation based unified model provides a good description of the $z$ $<$ 1 Seyfert population.  
The unified model does not seem to hold at $z$ $>$ 1, where quasar activity triggered by major mergers 
becomes more prevalent.

\textbullet Multi-wavelength data is consistent with the paradigm in which (high Eddington ratio) 
CT AGN have recently undergone intense star formation.  In the non-evolving 
model the AGN activity does not start until the starburst is $\gtrsim$1 Gyr old and 
the stellar population is highly obscured by dust.  In the evolving model, it is 
possible that the average high Eddington ratio CT AGN host galaxy is a LIRG 
or ULIRG, with the AGN activity starting one average $\gtrsim$0.3 Gyr after the starburst.

\textbullet $\gtrsim$5-15$\%$ of the type 1 and type 2 AGN population may have enhanced levels of star formation, 
with average SFR $\approx$ 20 M$_{\odot}$ yr$^{-1}$.  

\textbullet If the enhanced star formation sources are included, the evolving 
model CT AGN hosts are unaffected, but the non-evolving model CT AGN hosts are not consistent with the AGN 
evolution scheme and instead are consistent with the orientation based unified scheme.  Thus, if CT AGN are 
similar to type 2 AGN, then CT AGN host galaxies are similar to type 2 AGN 
host galaxies; however, if CT AGN are a distinct population of AGN in a special evolutionary stage, then the host 
galaxies of CT AGN are also in a special evolutionary stage.

\textbullet In order to fit the type 1 AGN and host B band luminosity function it is necessary for $f_2$ to evolve with $L_X$.  
However, future work studying the dependence of $f_2$ on redshift is necessary to elucidate the 
connection between AGN and their hosts.  Understanding the redshift evolution of $f_2$ will 
offer insight into how AGN are fueled and the nature of AGN obscuration.

\textbullet While the 
actual values of $E(B-V)$ depend on the metallicity and IMF used in the \citet{bc03} 
host galaxy population synthesis model, the qualitative findings of this study are independent 
of, and robust against the uncertainties in, the host galaxy metallicity and IMF.

\acknowledgments
The authors thank J. Donley and C. Pierce for comments on a draft of this paper.  This work was supported by NSF award AST 1008067.


{}


%
\begin{deluxetable}{ccccc}
\tablecolumns{5}
\tablecaption{Ages of stellar populations in Gyr for various redshift bins.}
\tablehead{
  \colhead{ } &
  \multicolumn{2}{c} {Type 1 and 2 AGN host} &
  \multicolumn{2}{c} {Low Eddington ratio CT AGN host\tablenotemark{\P}} \\
  \colhead{$z$} &
  \colhead{YSP\tablenotemark{\dag}} &
  \colhead{DSP\tablenotemark{\ddag}} &
  \colhead{YSP\tablenotemark{\dag}} &
  \colhead{DSP\tablenotemark{\ddag}} \\
  \colhead{} &
  \colhead{(0.1$M_{*}$)} &
  \colhead{(0.9$M_{*}$)} &
  \colhead{(0.1$M_{*}$)} &
  \colhead{(0.9$M_{*}$)} \\

}
\startdata
    $z$ $<$ 0.5 & 2.0 & 7.0 & 4.5 & 10 \\
    0.5 $<$ $z$ $<$ 1 & 2.0 & 7.0 & 4.5 & 7.0 \\
    1 $<$ $z$ $<$ 2 & 2.0 & 4.5 & 4.5 & 4.5 \\
    2 $<$ $z$ $<$ 3 & 2.0 & 2.0 & 2.0 & 2.0 \\
    3 $<$ $z$ & 1.0 & 1.0 & 1.0 & 1.0 \\
\enddata
\tablenotetext{\P}{The Low Eddington ratio CT AGN hosts are only relevant in the evolving model.  The age of the stellar population of the evolving model high Eddington ratio CT AGN hosts is the same in all $z$ bins and is shown in the bottom row of Table \ref{work}.}
\tablenotetext{\dag}{Younger Stellar Population}
\tablenotetext{\ddag}{Dominant Stellar Population}

\label{ages}
\end{deluxetable}
\begin{deluxetable}{lccccc}
\tablecolumns{6}
\tablecaption{$E(B-V)$ for the different IMFs and metallicities considered.  The first row summarizes the working model of this paper.}
\tablehead{
  \multicolumn{2}{c}{Model} & 
  \multicolumn{2}{c}{Unified/type 2 AGN host} & 
  \multicolumn{2}{c}{Type 1 AGN host} \\
  \colhead{IMF} &
  \colhead{Metallicity} &
  \colhead{$z$ $<$ 1} &
  \colhead{$z$ $>$ 1} &
  \colhead{$z$ $<$ 1} &
  \colhead{$z$ $>$ 1}
}
\startdata
    \citet{C03} & $Z_{\odot}$ & 0.25 & 0.5 & 0.25 & 0.05 \\
    \citet{C03} & 0.2$Z_{\odot}$ & 0.3 & 0.5 & 0.3 & 0.1 \\
    \citet{C03} & 2.5$Z_{\odot}$ & 0.2 & 0.7 & 0.1 & 0.05\\
    \citet{S55} & $Z_{\odot}$ & 0.1 & 0.1 & 0.1 & 0.0 \\
\enddata

\label{models}
\end{deluxetable}
\begin{deluxetable}{lcccc}
\tabletypesize{\small}
\tablecolumns{5}
\tablecaption{Summary of the AGN host stellar population parameters for the working model of this study.  The SFRs in parenthesis refer to the average SFR of the enhanced star formation sources.}
\tablehead{
  \colhead{AGN population} &
  \colhead{Stellar population age} &
  \colhead{$E(B-V)$} &
  \colhead{$E(B-V)$} &
  \colhead{SFR} \\
  \colhead{} &
  \colhead{(Gyr)} &
  \colhead{$z$ $<$ 1} &
  \colhead{$z$ $>$ 1} &
  \colhead{(M$_{\odot}$ yr$^{-1}$)} \\
}
\startdata
    Type 1 AGN & col 2 \& 3 Table \ref{ages} & 0.25 & 0.05 & 2.0 (20) \\
    Unified/type 2 AGN & col 2 \& 3 Table \ref{ages} & 0.25 & 0.5 & 2.0 (20) \\
    \hline \\
    {\bf Non-evolving model CT AGN} & & & & \\
    w/o enhanced star formation sources & 1.0 & 1.0 & 1.0 & 10 \\
    w/ enhanced star formation sources & col 2 \& 3 Table \ref{ages} & 0.25 & 0.5 & 2.0 (20) \\
    \hline \\
    {\bf Evolving model CT AGN} & & & & \\
    Low Eddington ratio CT AGN & col 4 \& 5 Table \ref{ages} & 1.0 & 1.0 & 1.0 \\
    High Eddington ratio CT AGN & 0.3 & 0.6 & 0.6 & 175 \\
\enddata

\label{work}
\end{deluxetable}
\clearpage
=
%
%
%
%
%
%
%
%

\end{document}